\documentclass[acmsmall,screen]{acmart}
\usepackage{amsmath,amsfonts}

\usepackage{amssymb}
\usepackage{amsthm}
\usepackage{graphicx}
\usepackage{subcaption}
\usepackage[table]{xcolor}
\usepackage{xspace}
\usepackage{booktabs}
\usepackage{multirow}
\usepackage[inline,shortlabels]{enumitem}
\usepackage{listings}
\usepackage{tcolorbox}
\tcbuselibrary{listings}

\usepackage{tikz}
\usetikzlibrary{arrows.meta,positioning,fit,calc,shapes.geometric}
\usepackage{pgfplots}
\usepgfplotslibrary{groupplots}
\pgfplotsset{compat=1.18}

\definecolor{cprimary}{HTML}{2166AC}    
\definecolor{csecondary}{HTML}{B2182B}  
\definecolor{ctertiary}{HTML}{1B7837}   
\definecolor{cquaternary}{HTML}{E08214} 
\definecolor{cfillA}{HTML}{D1E5F0}  
\definecolor{cfillB}{HTML}{D6EFC7}  
\definecolor{cfillC}{HTML}{FDDBC7}  
\definecolor{cfillD}{HTML}{F4A582}  
\definecolor{cneutral}{HTML}{F7F7F7}    
\definecolor{cborder}{HTML}{666666}      

\usepackage{algorithm}
\usepackage[noend]{algpseudocode}
\usepackage{subcaption}
\usepackage{threeparttable}
\usepackage{hyperref}

\newcommand{\baseline}{\textsc{Cobblestone}\xspace}
\newcommand{\tool}{\textsc{Quarry}\xspace}
\newcommand{\toolnohammer}{\textsc{Quarry-NoHammer}\xspace}
\newcommand{\toolnorank}{\textsc{Quarry-NoRank}\xspace}

\newcommand{\coqhammer}{CoqHammer\xspace}
\newcommand{\serapi}{SerAPI\xspace}

\newcommand{\coqgym}{CoqGym\xspace}
\newcommand{\coqgymtest}{CoqGym100\xspace}
\newcommand{\wigdersontest}{Wigderson100\xspace}
\newcommand{\wigderson}{coq-wigderson\xspace}
\newcommand{\cloverbenchtest}{TransBench58\xspace}

\newcommand{\gptfive}{GPT-5.2\xspace}

\newcommand{\mypara}[1]{\textbf{#1}}

\newsavebox{\coqlistingbox}

\definecolor{codegreen}{rgb}{0,0.6,0}
\definecolor{codepurple}{HTML}{8B008B}
\definecolor{codebg}{HTML}{F5F5F5}       
\definecolor{codeframe}{HTML}{CCCCCC}    
\lstdefinestyle{CoqStyle}{
  commentstyle=\color{codegreen}\itshape,
  keywordstyle=\color{cprimary}\bfseries,
  morekeywords={Require,Import,Export,Open,Scope,Section,End,Module,
    Proof,Qed,Admitted,Defined,Abort,
    Lemma,Theorem,Corollary,Proposition,Fact,Remark,
    Definition,Fixpoint,CoFixpoint,Function,Program,
    Inductive,CoInductive,Record,Structure,Class,Instance,
    Let,Variable,Variables,Hypothesis,Hypotheses,Parameter,Parameters,
    Notation,Infix,Reserved,Ltac,Tactic,
    Check,Print,Compute,Eval,Search,About,Set,Unset},
  keywordstyle=[2]\color{codepurple},
  morekeywords=[2]{intros,intro,apply,eapply,exact,eexact,
    rewrite,erewrite,simpl,cbv,cbn,unfold,fold,red,hnf,
    destruct,induction,case,case_eq,elim,
    split,left,right,exists,econstructor,constructor,
    assert,pose,pose proof,set,remember,generalize,specialize,
    clear,rename,move,revert,
    auto,eauto,tauto,omega,lia,nia,ring,field,
    trivial,easy,now,done,
    repeat,try,first,solve,progress,idtac,fail,
    reflexivity,symmetry,transitivity,congruence,subst,
    inversion,injection,discriminate,contradiction,exfalso,
    change,pattern,replace,cut,enough},
  keywordstyle=[3]\color{ctertiary},
  morekeywords=[3]{forall,exists,fun,match,with,end,
    if,then,else,let,in,return,as,where,
    Prop,Type,Set,SProp,
    nat,bool,list,option,unit,prod,sum,sig,sigT,
    true,false,Some,None,O,S,nil,cons,pair,tt,
    True,False,and,or,not,iff,eq},
  numberstyle=\tiny\color{cborder},
  stringstyle=\color{csecondary},
  basicstyle=\ttfamily\small,
  breakatwhitespace=false,
  breaklines=true,
  captionpos=b,
  keepspaces=true,
  numbers=left,
  xleftmargin=5.0ex,
  numbersep=5pt,
  showspaces=false,
  showstringspaces=false,
  showtabs=false,
  tabsize=2,
  literate={~}{{\raisebox{0.5ex}{\texttildelow}}}1 {<>}{{<>}}2
    {->}{{\textcolor{cborder}{$\rightarrow$}}}2
    {<>}{{\textcolor{cborder}{$\neq$}}}2
    {=>}{{\textcolor{cborder}{$\Rightarrow$}}}2
    {forall}{{\textcolor{ctertiary}{\textbf{forall}}}}6
}

\tcbset{
  coqbox/.style={
    colback=codebg,
    colframe=codeframe,
    boxrule=0.5pt,
    arc=2pt,
    left=2pt, right=2pt, top=2pt, bottom=2pt,
    fonttitle=\scriptsize\bfseries,
    coltitle=black,
  },
  coqbox-titled/.style 2 args={
    coqbox,
    title={#1},
    colbacktitle=codebg,
    coltitle=#2,
    fonttitle=\scriptsize\bfseries,
  },
}

\newtcolorbox{takeawaybox}{
  colback=cneutral,
  colframe=cborder,
  boxrule=0.4pt,
  arc=1.5pt,
  left=4pt, right=4pt, top=3pt, bottom=3pt,
  before skip=6pt,
  after skip=6pt,
}

\setcopyright{none}
\copyrightyear{2026}
\acmYear{2026}
\acmDOI{XXXXXXX.XXXXXXX}
\acmJournal{PACMPL}

\begin{document}

\title{Planning to Hammer: Difficulty-Aware Decomposition for Automating Rocq Proofs}

\author{Ning Zhang}
\affiliation{%
  \institution{Nanjing University}
  \city{Nanjing}
  \country{China}
}
\email{522023320204@smail.nju.edu.cn}

\author{Nongyu Di}
\affiliation{%
  \institution{Nanjing University}
  \city{Nanjing}
  \country{China}
}
\email{dny@smail.nju.edu.cn}

\author{Zenan Li}
\affiliation{%
  \institution{ETH Z\"{u}rich}
  \city{Zurich}
  \country{Switzerland}
}
\email{zenan.li@inf.ethz.ch}

\author{Yuan Yao}
\affiliation{%
  \institution{Nanjing University}
  \city{Nanjing}
  \country{China}
}
\email{y.yao@nju.edu.cn}

\author{Xiaoxing Ma}
\affiliation{%
  \institution{Nanjing University}
  \city{Nanjing}
  \country{China}
}
\email{xxm@nju.edu.cn}

\renewcommand{\shortauthors}{Zhang et al.}

\begin{abstract}

As AI-generated code proliferates, formal verification---particularly through interactive theorem provers such as Rocq and Isabelle---becomes increasingly important for ensuring software correctness. However, producing machine-checked proofs in such provers remains a bottleneck. Existing solutions bring complementary strengths to proof automation:
large language models (LLMs) can propose high-level proof strategies but lack local rigor; automated tactics such as CoqHammer can reliably discharge many local goals, but lack long-range planning capabilities.
To combine the best of both worlds, we present \tool, a \emph{planning-based} proof synthesis framework that separates proof planning from proof execution.
Specifically, \tool asks an LLM to actively propose multiple proof decompositions with arbitrary sublemmas, type-checks them in Rocq under temporarily admitted sublemmas, and ranks candidates using a proof-state-based difficulty model estimating hammer solvability.
It then recursively proves sublemmas within a bounded budget, effectively turning long proofs into sequences of hammer-solvable obligations.
We implement \tool on top of SerAPI and CoqHammer and evaluate it using multiple frontier LLMs across multiple benchmarks.
The experimental results show that planning-based decomposition with solvability-aware ranking substantially improves automation while maintaining predictable cost.
Under a uniform 10-minute wall-clock budget, \tool improves over the strongest baseline by 7\%--13\% in success rate across three Rocq benchmarks.
These results demonstrate that reliable proof automation can be achieved by coordinating neural planning with symbolic execution rather than replacing either.
\end{abstract}



\maketitle

\section{Introduction}                                     
\label{sec:intro}

Interactive theorem provers (ITPs) such as Rocq (formerly known as Coq)~\citep{coq} and  Isabelle/HOL~\citep{nipkow2002isabelle} provide a practical route to machine-checkable software assurance, and have been
used to verify compilers~\citep{Leroy09}, operating systems~\citep{Gu16}, and other critical artifacts~\citep{chen2015fscq,wilcox2015verdi,Erbsen19,Ileri18,li2023spoq}.
In the context of program verification, ITPs are used to prove that an implementation meets its formal specification.
The proof typically consists of a sequence of \emph{tactics} that transform the current \emph{proof state} until all obligations are discharged; the final proof object is then checked by the ITP kernel for formal assurance.
However, producing these proofs remains labor-intensive, as it requires reasoning about interacting program definitions, data-structure invariants, and specification predicates.
This demands significant expertise, limiting formal verification's adoption in current software practice.

In this paper, we study \emph{automated proof synthesis for program verification in Rocq\footnote{While we focus on Rocq in this work, the general idea is applicable to other ITPs.}}: given a program and its specification (e.g., a theorem about program correctness), along with its surrounding development
context (e.g., definitions and auxiliary lemmas), the task is to produce a complete, kernel-checked proof without manual intervention.
This problem is challenging because program-verification proofs must coordinate multiple forms of reasoning: unfolding nested definitions, establishing data-structure
  invariants, performing case analysis or induction over algebraic types, and connecting intermediate facts to specification predicates.
A single theorem may require dozens of tactic steps that interleave these reasoning modes, and small errors in any step invalidate the entire proof.
Moreover, the proof space is combinatorially large: each proof state admits many candidate tactics, and the search tree grows exponentially with proof length.

Existing approaches based on large language models (LLMs) or automated theorem provers (ATPs) each address part of the challenge, but neither alone is sufficient~\citep{li2024survey,yang2026formal}.
LLMs can suggest plausible proof structures but lack reliable local correctness, producing scripts in which small errors can derail the proof.
Conversely, hammer-style automation such as \coqhammer~\citep{Czajka18} can discharge many local goals by selecting relevant premises and dispatching them to external
ATPs, but it operates on a single goal at a time and struggles when the proof requires long-horizon reasoning, such as multiple intermediate steps, induction, or deep case analysis.

\mypara{Running example.}
We illustrate the gap with a graph-theory lemma from the \wigderson project~\citep{Phipathananunth23Wigderson}, which we revisit throughout the paper.
Specifically, the lemma \texttt{max\_deg\_remove\_node} states that removing a vertex from a graph preserves the maximum degree, under conditions on the degrees and adjacency of the involved nodes.
The proof of this lemma requires coordinating four interacting definitions (\texttt{degree}, \texttt{max\_deg}, \texttt{remove\_node}, and \texttt{adj}) and spans over 25 tactics in the human-written version, including nested case analysis and auxiliary assertions.
Neither \coqhammer nor frontier LLMs can solve it directly; we analyze the reasons in detail in Section~\ref{sec:motivation}.
In contrast, \tool's difficulty-aware planning decomposes it into 12 subgoals and solves it automatically.

\begin{tcolorbox}[
  colback=white,
  colframe=codeframe,
  boxrule=0pt,
  leftrule=1pt,
  arc=0pt,
  left=1pt, right=2pt, top=0pt, bottom=0pt,
  before skip=4pt,
  after skip=4pt,
]
\begin{lstlisting}[style=CoqStyle,numbers=left,xleftmargin=1.5em,basicstyle=\ttfamily\footnotesize,backgroundcolor=\color{white},aboveskip=0pt,belowskip=0pt]
Lemma max_deg_remove_node :
  forall (n : nat) (g : graph) (v x : node),
    degree v g = Some n ->
    degree x g = Some n ->
    max_deg g = n ->
    ~ S.In x (adj g v) ->
    x <> v ->
    max_deg (remove_node x g) = n.
\end{lstlisting}
\end{tcolorbox}

\mypara{Our work.}
To bridge this gap, we present \tool, a planning-based framework that separates \emph{proof planning} from \emph{proof execution}.
The core idea is to let an LLM decompose a theorem into multiple sublemmas along with a target proof script that closes the original goal assuming these sublemmas hold.
\coqhammer then discharges the sublemmas individually.
When a sublemma is too complex for direct automation, the system recursively applies the same decomposition, effectively turning a long proof into a tree of individually manageable obligations.

\mypara{The key challenge.}
An LLM can propose many candidate decompositions for a given goal, but not all decompositions are equally useful.
A decomposition is only as easy as its hardest sublemma: if even one sublemma exceeds the automation's reach, the entire recursive attempt fails and wastes the bounded budget.
Because the system must commit to a candidate before recursively solving its sublemmas (each of which involves hammer invocations and potentially further decomposition), the choice of which candidate to try first is critical.
A natural approach is to evaluate each candidate by actually running \coqhammer on all of its sublemmas, but this would be prohibitively expensive: with multiple candidates each containing several sublemmas and a 30-second hammer timeout per sublemma, the evaluation cost alone could exhaust the time budget.
We therefore need a lightweight \emph{difficulty proxy} that estimates sublemma solvability from proof-state features alone, without invoking the hammer.

\mypara{Our key insight.}
We observe that in the program-verification setting, the proving difficulty of a sublemma correlates with measurable features of its proof state, including the computational complexity of the referenced definitions and the logical complexity of the proof obligation.
By scoring candidate decompositions according to these features, the system can prioritize plans whose sublemmas are predicted to be solvable, spending its limited budget on the most promising proof strategies.
Furthermore, this scoring is flexible: by assigning weights to different proof-state features, the difficulty model can adapt to the strengths and limitations of the underlying automation.

\mypara{System: \tool.}
We operationalize these ideas in \tool, a system built around a \emph{Generate--Rank--Solve} loop (Section~\ref{sec:approach}).
Given a proof obligation, the LLM proposes multiple candidate decompositions, which are verified against Rocq's type-checker, ranked by a learned difficulty model, and recursively solved using \coqhammer as the primary execution engine.
The entire process operates within a bounded budget that ensures predictable cost.

We evaluate \tool on benchmarks spanning diverse Rocq developments, graph-theory verification, and Rust-to-Rocq translations, using multiple LLM backends.
\tool proves 55\%, 52\%, and 16\% of theorems on the three benchmarks respectively, improving over the strongest baseline by 7--13 percentage points in success rate under a uniform 10-minute wall-clock budget.

\mypara{Contributions.}
This paper makes the following contributions:
\begin{itemize}[leftmargin=*]
  \item A formulation of Rocq proof synthesis as a planning-then-execution problem, where an LLM proposes arbitrary sublemma decompositions and \coqhammer discharges the resulting obligations.
  \item \tool, a \emph{Generate--Rank--Solve} framework that combines Rocq-side candidate verification, difficulty-aware ranking, and hammer-first recursive solving under a bounded budget.
  \item An evaluation across three benchmarks and multiple LLM backends under a uniform 10-minute wall-clock budget, showing 7\%--13\% improvements over the strongest baseline.
\end{itemize}

\mypara{Roadmap.}
The rest of the paper provides background knowledge (Section~\ref{sec:background}) and analyzes a motivating example (Section~\ref{sec:motivation}), presents \tool's framework
(Section~\ref{sec:approach}) and its difficulty-aware ranking component (Section~\ref{sec:ranking}), reports the evaluation results (Section~\ref{sec:eval}), discusses threats to validity and related work
(Sections~\ref{sec:threats}--\ref{sec:related}), and concludes (Section~\ref{sec:conclusion}).

\section{Background}
\label{sec:background}

This section introduces the basic concepts used throughout the paper.
We first describe the Rocq proof assistant as well as its proof language, and then introduce \coqhammer and \serapi, the two infrastructure components that \tool builds upon.

\subsection{Rocq and its Proof Language}
\label{sec:bg-coq}

Rocq~\citep{coq} (historically known as Coq) is an interactive theorem prover based on the Calculus of Inductive Constructions (CIC).
It has been used to verify various software artifacts such as the CompCert compiler~\citep{Leroy09} and the CertiKOS operating system kernel~\citep{Gu16}.

In Rocq, a user can state a \emph{theorem} (or \emph{lemma}) as a logical proposition to be proved. This proof goal can be further decomposed into several subgoals (or proof obligations). Then, a machine-checked \emph{proof}, which is usually written as a sequence of \emph{tactics}, can be constructed to prove each subgoal.
A theorem is proved once all its subgoals have been resolved.
\begin{itemize}[leftmargin=*,itemsep=0.3ex]
  \item {Proof state}: A {\em proof state} consists of a goal (the proposition to be proved) together with a \emph{context} of available hypotheses and previously established facts. We write $\Gamma \vdash G$ for a proof state with context $\Gamma$ and goal $G$.
  \item {Tactic}: A {\em tactic} is a command that transforms the current proof state. Common tactics include:
  \begin{itemize}[itemsep=0ex]
    \item \texttt{intros}, which moves universally quantified variables and hypotheses from the goal into the context.
    \item \texttt{split}, \texttt{destruct}, \texttt{induction}, which decompose a goal into multiple {subgoals} (e.g., splitting a conjunction $A \wedge B$ yields subgoals $A$ and $B$).
    \item \texttt{apply}, \texttt{rewrite}, \texttt{unfold}, which reshape the goal without necessarily creating subgoals (e.g., applying a known lemma or unfolding a definition).
  \end{itemize}
  \item {Proof script}: A {\em proof script} is the complete tactic sequence that transforms the initial proof state into one with no remaining subgoals, enclosed between \texttt{Proof} and \texttt{Qed}.
\end{itemize}

\subsection{CoqHammer}
\label{sec:bg-hammer}

\coqhammer~\citep{Czajka18} bridges Rocq with external automated theorem provers (ATPs) such as satisfiability modulo theories (SMT) solvers.
Given a proof obligation, it operates in two phases:
\begin{enumerate}[leftmargin=*,itemsep=0.3ex]
  \item {Premise selection}: analyze the current goal and context to identify a relevant subset of previously established lemmas and definitions.
  \item {ATP invocation and reconstruction}: translate the goal and selected premises into first-order logic, dispatch the problem to external provers (e.g., E, Vampire, Z3), and, if a prover succeeds, reconstruct a Rocq-checkable proof.
\end{enumerate}
Since the reconstructed proof is verified by the Rocq kernel, \coqhammer provides the same soundness guarantee as a manually written proof.

\coqhammer excels at goals involving equational reasoning, set-theoretic manipulations, or combinations of known facts.
However, it struggles with goals that require induction, deep case analysis, or long chains of intermediate steps.
In our running example, \coqhammer alone cannot prove \texttt{max\_deg\_remove\_node}, as the proof requires reasoning about multiple interacting definitions and a non-trivial case split.
Yet, as we will show, \coqhammer \emph{can} discharge many of the smaller subgoals that arise once the goal is decomposed.

\subsection{SerAPI}
\label{sec:bg-serapi}

\serapi~\citep{GallegoArias16} exposes a programmatic interface to the Rocq proof engine, allowing external tools to interact with it directly.
Through \serapi, a client can:
\begin{itemize}[leftmargin=*,itemsep=0.3ex]
  \item \emph{Step through proofs} by submitting individual tactics and observing the resulting proof state;
  \item \emph{Inspect goals and hypotheses} at each step, including their full abstract syntax trees (ASTs);
  \item \emph{Collect error feedback} when a tactic fails, enabling automated retry or recovery.
\end{itemize}
\tool relies on \serapi for three purposes: (1)~\emph{validating} candidate decompositions by executing proposed tactic scripts and checking that they type-check, (2)~\emph{extracting} intermediate proof states for difficulty estimation (Section~\ref{sec:rank}), and (3)~\emph{recovering} from transient failures (e.g., timeouts) during the recursive search.

\section{Motivation}
\label{sec:motivation}
We use the \texttt{max\_deg\_remove\_node} lemma introduced in Section~\ref{sec:intro} to illustrate the challenges of automated proof synthesis.
This lemma is representative of program-verification theorems that require coordinating multiple definitions and intermediate facts, where neither a single hammer invocation nor a single LLM generation suffices.
We first show \emph{why} existing approaches fail (Sections~\ref{sec:mot-hammer}--\ref{sec:mot-llm}), analyze the gap (Section~\ref{sec:gap}), and finally preview the decomposition that \tool finds (Section~\ref{sec:preview}).

\subsection{Why a Single Hammer Call Fails}
\label{sec:mot-hammer}

We invoke \coqhammer on the original goal of the running example:
\begin{tcolorbox}[
  colback=white,
  colframe=codeframe,
  boxrule=0pt,
  leftrule=1pt,
  arc=0pt,
  left=1pt, right=2pt, top=0pt, bottom=0pt,
  before skip=4pt,
  after skip=4pt,
]
\begin{lstlisting}[style=CoqStyle,numbers=left,xleftmargin=1.5em,basicstyle=\ttfamily\footnotesize,backgroundcolor=\color{white},aboveskip=0pt,belowskip=0pt]
forall (n : nat) (g : graph) (v x : node),
  degree v g = Some n -> degree x g = Some n ->
  max_deg g = n -> ~ S.In x (adj g v) ->
  x <> v -> max_deg (remove_node x g) = n.
\end{lstlisting}
\end{tcolorbox}
After introducing the hypotheses, the proof state contains five hypotheses and a goal involving \texttt{max\_deg} applied to \texttt{remove\_node}.
To close this goal, \coqhammer must simultaneously reason about graph structure, degree computation, adjacency sets, and node removal.
The proof requires establishing both directions of an equality via antisymmetry ($\le$ and $\ge$), each of which depends on intermediate properties not directly available as premises.
All external ATPs time out: the search space, spanning multiple definitions and their interactions, is too large for a single query.

\subsection{Why a Single LLM Generation Fails}
\label{sec:mot-llm}

We ask one of the most advanced LLMs (i.e., \gptfive) to generate a complete proof script for the running example. 
We sampled 8 independent candidates; all were rejected by Rocq.
The candidates are remarkably consistent in high-level strategy: all eight candidates split the proof into two directions (upper bound $\texttt{max\_deg (remove\_node x g)} \le n$ and lower bound $n \le \texttt{max\_deg (remove\_node x g)}$) and attempt to close via antisymmetry, matching the structure of the human proof.

The LLM thus demonstrates sound high-level planning: it correctly identifies the proof skeleton (antisymmetry, then prove each direction).
However, realizing this plan requires ${\sim}10$ auxiliary lemmas totaling ${\sim}$176 lines of Rocq, and the model cannot expand all of these low-level details coherently within a single \texttt{Proof\ldots Qed} block.
This gap between correct planning and incomplete execution manifests in three ways:

\begin{enumerate}[leftmargin=*,itemsep=0ex]
\item \emph{Hallucinated auxiliary lemmas.}
  Each candidate invokes 2--5 non-existent lemmas (for example, \texttt{max\_deg\_non\_increasing}).
  The model identifies \emph{what} intermediate facts are needed, but cannot define and prove them within a single generation.

\item \emph{Incomplete or abandoned subproofs.}
  When the model recognizes that a subgoal exceeds its generation capacity, it resorts to \texttt{Admitted} or leaves holes, producing proof scripts that compile only under trust assumptions and are rejected by the Rocq kernel.

\item \emph{Incorrect tactic chains.}
  Even when the model attempts to fill in low-level details, it generates ill-typed tactic sequences.
  A representative error: in this project, the notation \texttt{<=} resolves to \texttt{Pos.le} (positive integers) rather than \texttt{Nat.le};
  seven of eight candidates write \texttt{max\_deg \ldots <= max\_deg \ldots} without a \texttt{\%nat} scope annotation, triggering an immediate type mismatch.
\end{enumerate}

In summary, the LLM \emph{can} identify the right proof plan, but a single-pass generation cannot simultaneously produce all the auxiliary machinery that plan requires.
Generating all required pieces coherently in one LLM call amounts to searching an $O(b^d)$ space (branching factor $b$, logical depth $d$) in a single step---a fundamentally intractable task for current models.

\subsection{Analyzing the Gap}
\label{sec:gap}

The two approaches fail for complementary reasons: the hammer lacks planning capability, while the LLM lacks execution reliability.
The core difficulty is that this theorem has a \emph{long proof} (over 25 tactics in the human version) involving \emph{five interacting definitions}, yet the proof structure is not visible from the goal statement alone.
We identify four structural requirements for a successful approach:

\mypara{Requirement 1: planning, not just execution.}
The hammer executes but cannot plan: it tries to solve the current goal in one shot.
Single-pass LLM generation (as in Section~\ref{sec:mot-llm}) produces a full tactic sequence without feedback, so errors compound across steps.
Even step-by-step approaches that interact with the prover~\citep{Ringer21a,Blaauwbroek20} still operate at the tactic level without an explicit high-level plan.
A successful approach should \emph{plan} the proof structure, breaking the goal into manageable pieces, before attempting to execute each piece.

\mypara{Requirement 2: arbitrary intermediate lemmas.}
\baseline~\citep{kasibatla2026cobblestone}, the state-of-the-art LLM-based proof-repair method, generates whole proof scripts, localizes errors via fail-safe execution, and recursively retries on failing subproofs.
Its decomposition is driven by Rocq's built-in subgoal mechanism: it keeps working subproofs and re-generates only the failing ones, so the sub-obligations it creates are limited to those that arise from tactic execution (e.g., \texttt{split}, \texttt{induction}).
However, this theorem requires an \texttt{apply Nat.le\_antisymm} followed by \texttt{assert} calls that introduce custom intermediate lemmas, steps that do not naturally produce subgoals through prefix execution.
\baseline's own analysis (RQ5) confirms this: proofs that make ``internal progress'' without creating explicit subgoals are a systematic failure mode of its error-driven repair strategy.
A successful approach must therefore be able to introduce \emph{arbitrary} intermediate lemmas, not just those that arise from Rocq's built-in goal-splitting.

\mypara{Requirement 3: difficulty-aware selection.}
Even if a system can propose decompositions, not all decompositions are equally useful.
Many plausible-looking decompositions contain a ``bottleneck'' sublemma that is too hard for automation, causing the entire recursion to fail.
A successful approach must be able to \emph{evaluate} the difficulty of proposed sublemmas and prioritize decompositions where all sublemmas are within the automation's reach.

\mypara{Requirement 4: predictable cost.}
Proof synthesis is inherently expensive: each LLM call and each hammer invocation consumes time and computational resources.
Iterative repair approaches such as \baseline can enter long retry loops, where each failure triggers a full re-generation, leading to unpredictable and often excessive wall-clock cost.
A successful approach must operate within a \emph{bounded budget}, ensuring that the total cost remains predictable regardless of whether the proof attempt succeeds or fails.

\subsection{Preview: What a Successful Decomposition Looks Like}
\label{sec:preview}
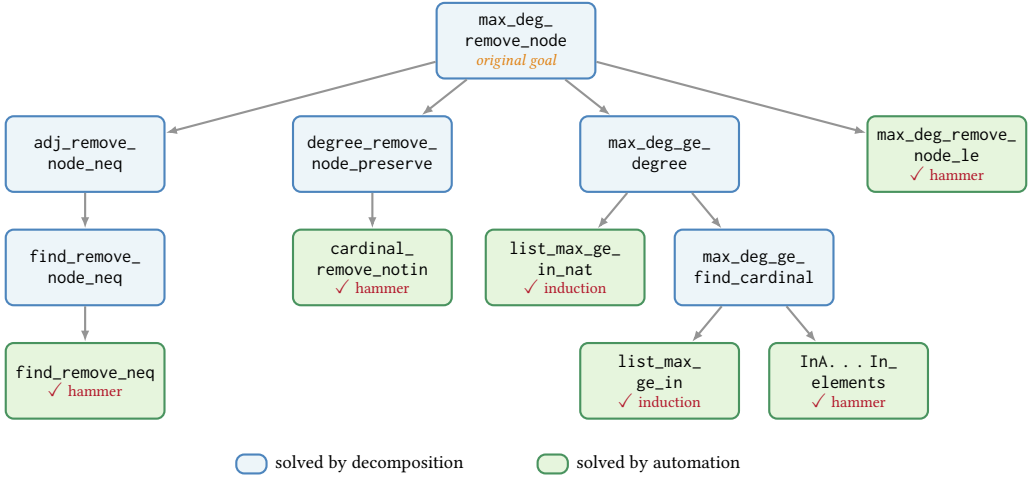
\begin{figure}[t]
\centering
\footnotesize
\begin{tikzpicture}[
  every node/.style={align=center, font=\scriptsize},
  level 1/.style={sibling distance=38mm, level distance=15mm},
  level 2/.style={sibling distance=25mm, level distance=15mm},
  level 3/.style={sibling distance=25mm, level distance=15mm},
  edge from parent/.style={draw=cborder!70, thick, -{Latex[length=1.5mm]}},
  decomp/.style={draw=cprimary!80, thick, rounded corners=3pt, fill=cfillA!40,
    inner sep=3pt, minimum width=21mm, minimum height=10mm,
    font=\scriptsize},
  solved/.style={draw=ctertiary!80, thick, rounded corners=3pt, fill=cfillB!60,
    inner sep=3pt, minimum width=21mm, minimum height=10mm,
    font=\scriptsize},
]
\node[decomp] {\texttt{max\_deg\_}\\{}\texttt{remove\_node}\\[-1pt]{\color{cquaternary}\tiny\itshape original goal}}
  child { node[decomp] {\texttt{adj\_remove\_}\\{}\texttt{node\_neq}}
    child { node[decomp] {\texttt{find\_remove\_}\\{}\texttt{node\_neq}}
      child { node[solved] {\texttt{find\_remove\_neq}\\[-1pt]{\color{csecondary}\checkmark\ {\tiny hammer}}} }
    }
  }
  child { node[decomp] {\texttt{degree\_remove\_}\\{}\texttt{node\_preserve}}
    child { node[solved] {\texttt{cardinal\_}\\{}\texttt{remove\_notin}\\[-1pt]{\color{csecondary}\checkmark\ {\tiny hammer}}} }
  }
  child { node[decomp] {\texttt{max\_deg\_ge\_}\\{}\texttt{degree}}
    child { node[solved] {\texttt{list\_max\_ge\_}\\{}\texttt{in\_nat}\\[-1pt]{\color{csecondary}\checkmark\ {\tiny induction}}} }
    child { node[decomp] {\texttt{max\_deg\_ge\_}\\{}\texttt{find\_cardinal}}
      child { node[solved] {\texttt{list\_max\_}\\{}\texttt{ge\_in}\\[-1pt]{\color{csecondary}\checkmark\ {\tiny induction}}} }
      child { node[solved] {\texttt{InA\ldots In\_}\\{}\texttt{elements}\\[-1pt]{\color{csecondary}\checkmark\ {\tiny hammer}}} }
    }
  }
  child { node[solved] {\texttt{max\_deg\_remove\_}\\{}\texttt{node\_le}\\[-1pt]{\color{csecondary}\checkmark\ {\tiny hammer}}} };

\node[decomp, thick, minimum width=4mm, minimum height=2.5mm, inner sep=0pt,
  label={[font=\scriptsize, anchor=west]right:solved by decomposition}]
  (leg1) at (-3.5, -5.6) {};
\node[solved, thick, minimum width=4mm, minimum height=2.5mm, inner sep=0pt,
  label={[font=\scriptsize, anchor=west]right:solved by automation}]
  at ([xshift=42mm]leg1.west) {};
\end{tikzpicture}
\caption{Proof tree for \texttt{max\_deg\_remove\_node} found by \tool.
The tree has depth~3 and 12 nodes. Green leaves are discharged by \coqhammer or simple induction; blue nodes required further decomposition.
Each leaf involves only one or two definitions, making it tractable for automation.}
\label{fig:motivation-tree}
\end{figure}

Figure~\ref{fig:motivation-tree} shows the decomposition that \tool finds for this theorem.
The key insight is that the equality \texttt{max\_deg (remove\_node x g) = n} can be split via antisymmetry into two inequality directions, each supported by dedicated sublemmas.
The root-level decomposition uses \texttt{Nat.le\_antisymm} to split the equality, then introduces four sublemmas:
\begin{itemize}[leftmargin=*,itemsep=0ex]
  \item \texttt{max\_deg\_remove\_node\_le}: the max degree of a subgraph does not exceed that of the parent graph (directly solved by \coqhammer);
  \item \texttt{degree\_remove\_node\_preserve}: removing a non-adjacent node preserves the degree of $v$ (requires one further sublemma about set cardinality);
  \item \texttt{max\_deg\_ge\_degree}: the max degree is at least as large as any individual node's degree (requires two recursive sublemmas about list maxima);
  \item \texttt{adj\_remove\_node\_neq}: the adjacency set after node removal can be characterized in terms of the original adjacency set (requires reasoning about map lookups).
\end{itemize}

This decomposition addresses all four requirements: it plans a tree structure before executing any leaf (Req.~1), introduces arbitrary sublemmas beyond Rocq's built-in subgoal mechanism (Req.~2), selects among candidate decompositions based on estimated sublemma difficulty (Req.~3), and completes within a bounded budget (Req.~4).
Crucially, each sublemma involves only one or two definitions---compared to the four interacting definitions in the original goal---reducing the effective search depth per subgoal dramatically (we formalize this reduction in Section~\ref{sec:decomposition-quality}).

\section{Approach}
\label{sec:approach}
In this section, we present the proposed \tool. We start with its overview, and then describe how \tool systematically generates, evaluates, and selects the decompositions.

\subsection{Overview}
\label{sec:overview}

\begin{figure*}[t]
\centering
\includegraphics[width=\textwidth]{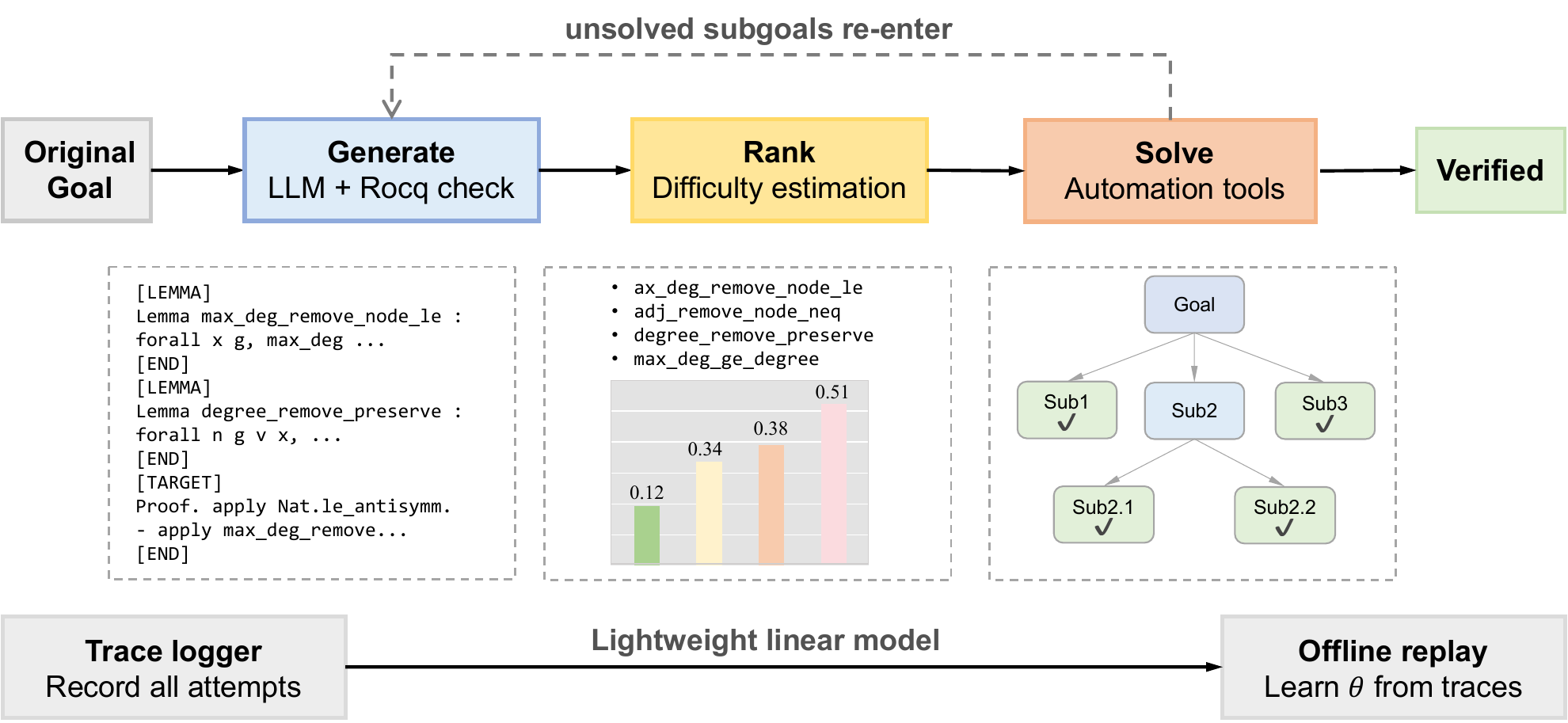}
\caption{\tool workflow.
Given a goal, the system enters the Generate--Rank--Solve loop:
the LLM proposes $k$ candidate decompositions, which are filtered by Rocq-side verification (\texttt{Admitted} sublemmas);
the difficulty model ranks survivors by estimated solvability;
and the top-$B$ candidates are recursively solved using automation tools (e.g., \coqhammer).
Unsolved subgoals re-enter the loop (dashed arrow).
The bottom row shows the offline learning pipeline: a trace logger records all attempts, and offline replay learns the difficulty model weights $\theta$.}
\label{fig:Quarry-workflow}
\end{figure*}

Rather than generating whole proofs and reactively fixing errors, \tool is a \emph{planning-based} proof synthesis framework that generates candidate decompositions, evaluates their feasibility, and selects the most promising candidate before committing to any sub-proof.
Specifically, given a proof state $\Gamma \vdash G$, \tool interacts with Rocq via \serapi to solve $G$ under a fixed computational budget.
As illustrated in Fig.~\ref{fig:Quarry-workflow}, \tool repeatedly applies three phases:
\begin{enumerate}[leftmargin=*,itemsep=0.3ex]
  \item \textbf{Generate}: \tool samples multiple candidate decompositions from an LLM, each proposing a set of sublemmas and a target proof script, and filters out ill-typed or circular candidates via Rocq-side verification (Section~\ref{sec:generate-verify}).
  \item \textbf{Rank}: The surviving candidates are scored by a proof-state-based difficulty model that estimates automation solvability, and ranked by the maximum difficulty among their sublemmas (Section~\ref{sec:ranking}).
  \item \textbf{Solve}: \tool recursively proves sublemmas of the top-ranked candidate via the same loop, using the automation backend as a fast path whenever possible (Section~\ref{sec:solve}).
\end{enumerate}

The Generate--Rank--Solve loop is motivated by a key observation: when the automation backend (e.g., \coqhammer\footnote{The framework is agnostic to the choice of automation backend; we instantiate it with \coqhammer in our implementation and evaluation.}) cannot solve a goal directly, the system generates multiple candidate decompositions, but \emph{only the selected candidate's sublemmas enter the recursive solving process}.
Recursively solving a candidate is expensive---each sublemma triggers its own hammer attempt and potentially further LLM-based decomposition---so the system cannot afford to try all candidates exhaustively.
By default, \tool commits to a single top-ranked candidate per node ($B=1$), making candidate selection a critical decision. 
Difficulty-aware ranking addresses this by estimating sublemma solvability from lightweight proof-state features \emph{before} committing to any candidate, directing the budget toward the decomposition most likely to succeed.

The remainder of this section details the Generate and Solve phases; Algorithm~\ref{alg:main} outlines the complete online search process.
Since the ranking component involves its own feature-engineering, model-training, and offline-learning pipeline---distinct in nature from the system-level Generate and Solve machinery---we present it separately in Section~\ref{sec:ranking}.

\subsection{Generate: Decomposition and Verification}
\label{sec:generate-verify}

Instead of predicting a monolithic proof script, \tool asks the LLM to propose candidate decompositions.
At a proof node $n$ with goal $G_n$, the LLM generates up to $k$ raw candidates (controlled by the rollout budget).
Each candidate decomposition $c$ consists of:
\begin{enumerate}[(1), leftmargin=*, itemsep=0ex]
    \item \emph{Sublemmas} $L(c) = \{\ell_1, \dots, \ell_m\}$, where each $\ell_i$ is a Rocq lemma statement;
    \item \emph{Target proof} $p(c)$, a tactic script intended to solve $G_n$ assuming $L(c)$ holds.
\end{enumerate}

If a candidate is selected and all its sublemmas are recursively proved, the target proof $p(c)$ is executed to close $G_n$; otherwise the system moves on to any remaining candidate in the top-$B$ set.
With the default $B=1$, this means that no alternative decomposition is tried at that node.
In our implementation, the LLM is instructed to emit candidates in a structured, parseable format: a sequence of \texttt{[LEMMA]}\,/\texttt{[END]} blocks and exactly one \texttt{[TARGET]}\,/\texttt{[END]} block.
Any response that cannot be parsed into this schema is treated as an invalid candidate.

To prevent the search space from exploding with ill-formed or unsound decompositions, \tool implements a strict Rocq-based verifier.
The parent goal is considered \emph{conditionally closed} when the candidate proof script type-checks in Rocq assuming the candidate sublemmas are \texttt{Admitted}.
A candidate $c$ is considered \emph{valid} if and only if:
\begin{enumerate}[(a), leftmargin=*, itemsep=0ex]
    \item All sublemmas $\ell_i \in L(c)$ can be parsed and accepted by Rocq using the \texttt{Admitted} vernacular;
    \item The proposed target proof $p(c)$ successfully type-checks and closes the goal $G_n$ under the context augmented by the admitted $\ell_i$.
\end{enumerate}
Candidates failing this semantic check, as well as trivial circularities (e.g., $\ell_i$ is syntactically identical to $G_n$), are immediately discarded. The surviving candidates form the \emph{valid candidate set} $V$.

To make the Generate phase concrete, Figure~\ref{fig:decomposition-example} shows how \tool processes the root goal of our running example.
The LLM proposes a decomposition with four sublemmas and a target proof that uses \texttt{Nat.le\_antisymm} to split the equality into two inequality directions.
The verifier checks that the target proof type-checks when the four sublemmas are \texttt{Admitted}, confirming conditions (a) and (b) above.
The difficulty model then scores each sublemma: \texttt{max\_deg\_remove\_node\_le} receives the lowest difficulty score (it is a direct consequence of the subgraph property), while \texttt{max\_deg\_ge\_degree} receives the highest (it requires further decomposition into lemmas about list maxima).
Based on this ranking, \tool prioritizes the decomposition and proceeds to recursively solve each sublemma.

\begin{figure}[t]
\centering
\footnotesize
\begin{tcolorbox}[
  colback=white,
  colframe=cprimary!80,
  boxrule=0pt,
  leftrule=2pt,
  arc=0pt,
  left=2pt, right=2pt, top=2pt, bottom=2pt,
  before skip=4pt,
  after skip=2pt,
]
{\scriptsize\bfseries\color{cprimary} Generate: LLM proposes decomposition for \texttt{max\_deg\_remove\_node}}
\begin{lstlisting}[style=CoqStyle,numbers=none,xleftmargin=0pt,basicstyle=\ttfamily\scriptsize,backgroundcolor=\color{white},aboveskip=2pt,belowskip=0pt]
[LEMMA]
Lemma max_deg_remove_node_le : forall x g,
  max_deg (remove_node x g) <= max_deg g.
[END]
[LEMMA]
Lemma degree_remove_node_preserve : forall n g v x,
  degree v g = Some n -> ~ S.In x (adj g v) ->
  x <> v -> degree v (remove_node x g) = Some n.
[END]
[LEMMA]
Lemma max_deg_ge_degree : forall g v n,
  degree v g = Some n -> max_deg g >= n.
[END]
... (* adj_remove_node_neq omitted for space *)
[TARGET]
Proof.
  repeat intro. apply Nat.le_antisymm.
  - rewrite <- H1. apply max_deg_remove_node_le.
  - eapply le_trans. 2: eapply max_deg_ge_degree.
    eapply degree_remove_node_preserve; eauto. eauto.
[END]
\end{lstlisting}
\end{tcolorbox}

\vspace{-1mm}
\begin{tcolorbox}[
  colback=white,
  colframe=ctertiary!80,
  boxrule=0pt,
  leftrule=2pt,
  arc=0pt,
  left=2pt, right=2pt, top=2pt, bottom=2pt,
  before skip=2pt,
  after skip=2pt,
]
{\scriptsize\bfseries\color{ctertiary} Verify: Rocq checks target proof with sublemmas \texttt{Admitted}}\\[1pt]
\scriptsize Target proof type-checks: \checkmark\quad
No circularity detected: \checkmark\quad
Valid candidate added to $V$.
\end{tcolorbox}

\vspace{-1mm}
\begin{tcolorbox}[
  colback=white,
  colframe=csecondary!80,
  boxrule=0pt,
  leftrule=2pt,
  arc=0pt,
  left=2pt, right=2pt, top=2pt, bottom=2pt,
  before skip=2pt,
  after skip=2pt,
]
{\scriptsize\bfseries\color{csecondary} Rank: difficulty model scores sublemmas}\\[1pt]
\scriptsize
\begin{tabular}{@{}lcc@{}}
Sublemma & $d_\theta(\ell)$ & Predicted solvability \\
\texttt{max\_deg\_remove\_node\_le} & 0.12 & \coqhammer (direct) \\
\texttt{degree\_remove\_node\_preserve} & 0.38 & further decomposition \\
\texttt{max\_deg\_ge\_degree} & 0.51 & further decomposition \\
\texttt{adj\_remove\_node\_neq} & 0.34 & further decomposition \\
\end{tabular}
\\[2pt]
Candidate score $= \max\{0.12, 0.38, 0.51, 0.34\} = 0.51$
\end{tcolorbox}
\caption{How \tool processes the root goal of the running example through the Generate--Rank pipeline.
The LLM produces a structured decomposition; Rocq verification (part of Generate) confirms conditional correctness; the difficulty model estimates sublemma solvability.}
\label{fig:decomposition-example}
\end{figure}

\subsection{Solve: Online Search and Execution}
\label{sec:solve}

With the ranked candidates $V_{\text{ranked}}$, \tool performs a depth-first recursive search (Algorithm~\ref{alg:main}).
For each node, it first attempts the automation backend (in our case, \coqhammer) as a cheap, trusted fast path; if it fails, the system generates, verifies, and ranks decompositions as described above.
The online solver then commits to the top-$B$ ranked candidates per node ($B=1$ by default).
With $B=1$, each node gets exactly one attempt: if the top-ranked candidate's sublemmas cannot all be recursively proved, the node is marked as failed without trying alternative decompositions.

For each selected candidate $c$, \tool recursively solves its sublemmas in the order generated by the LLM.
If all sublemmas are proved, the target proof $p(c)$ is executed to close the parent goal and the node is marked as solved.
The runner maintains a shared proven-context of discharged sublemmas: when the same sublemma appears later in the recursive search, \tool reuses the existing proof instead of proving it again.
Importantly, \tool uses \texttt{Admitted} only inside the verifier to test conditional solvability; a goal is returned as solved only after Rocq has accepted a proof script in which all previously admitted sublemmas have been discharged.
Thus, the final output is always a standard Rocq proof object checked by the kernel, and LLM outputs are treated as untrusted suggestions.
To gather denser supervision data for offline learning, \tool supports a \emph{dense-supervision mode} where it continues to attempt all remaining candidates at each node even after a success, recording every candidate's recursive outcome in the trace; otherwise, it returns early upon the first successful decomposition.
The dense-supervision traces are used for offline replay (Section~\ref{sec:offline}) and for the RQ5 evaluation (Section~\ref{sec:rq5}), as they provide the complete candidate--outcome mapping needed to evaluate alternative ranking policies without new LLM calls.

\renewcommand{\algorithmicrequire}{\textbf{Input:}}
\renewcommand{\algorithmicensure}{\textbf{Output:}}
\begin{algorithm}[ht]
\caption{\tool's Online Recursive Search.}
\label{alg:main}
\small
\begin{algorithmic}[1]
\Require Goal $G$; recursion depth \texttt{budget}; rollout budget $k$; top-candidate budget $B$; global LLM request limit (Section~\ref{sec:setup})
\Ensure \textbf{true} if $G$ is solved with a kernel-checked Rocq proof, \textbf{false} otherwise

\Function{SolveGoal}{$G, \text{budget}$}
  \State \textbf{// Phase 1: Fast path}
  \State \textbf{if} \Call{AutoSolver}{$G$} \textbf{then return true} \Comment{try automation, e.g., \coqhammer}
  \State \textbf{if} budget $\le 0$ \textbf{then return false}

  \State \textbf{// Phase 2: Generate and rank}
  \State $C \gets$ \Call{LLM\_Generate}{$G, k$} \Comment{sample $k$ candidates}
  \State $V \gets \{ c \in C \mid \Call{RocqVerify}{G, c} \}$ \Comment{filter via \texttt{Admitted}}
  \For{each $c \in V$, each $\ell \in L(c)$}
    \State $d_\theta(\ell) \gets \theta^\top \phi(\ell)$ \Comment{difficulty estimation}
  \EndFor
  \State $V_{\text{ranked}} \gets \Call{SortBy}{V,\; c \mapsto \max_{\ell \in L(c)} d_\theta(\ell),\; \text{ascending}}$
  \State $V_{\text{try}} \gets V_{\text{ranked}}[1..B]$ \Comment{top-$B$ candidates ($B{=}1$ by default)}

  \State \textbf{// Phase 3: Commit and recurse}
  \For{each candidate $c \in V_{\text{try}}$}
    \If{$\forall \ell \in L(c): \Call{SolveGoal}{\ell, \text{budget} - 1}$}
      \State \Call{ExecuteProof}{$p(c), G$} \Comment{close $G$ in Rocq}
      \State \Return \textbf{true}
    \EndIf
  \EndFor
  \State \Return \textbf{false}
\EndFunction
\end{algorithmic}
\end{algorithm}

\section{Difficulty-Aware Ranking and Learning}
\label{sec:ranking}

Recall that when \coqhammer cannot solve a goal directly, the Generate phase produces a set of valid candidate decompositions, each proposing different sublemmas.
Before committing to any candidate and spending the recursive budget on its sublemmas, the system must decide which candidate to try---and with the default $B=1$, which candidate to try \emph{at all}.
Therefore, discovering a successful decomposition earlier avoids expanding deep, fruitless recursive branches.

\subsection{Decomposition Quality and Search Cost}
\label{sec:decomposition-quality}

In a simplified tactic-level proof-search model, the prover explores a tree of candidate tactics: at each step it faces $b > 1$ alternatives, so a goal requiring $L$ steps costs $\Theta(b^L)$.
If the goal is decomposed into $r$ independent subgoals with depths $L_1, \dots, L_r$ summing to $L$, the total cost becomes $\sum_i b^{L_i}$ rather than $b^L$.
By the convexity of $f(x) = b^x$, this sum is minimized when all subgoals have equal depth $L_i = L/r$, giving $\Theta(r \cdot b^{L/r})$ in this model.
Hammer-style solvers such as \coqhammer use a different internal search strategy (translating goals to first-order logic and invoking external ATPs), but the same principle applies: their solving cost grows superlinearly with goal complexity, so decomposing a hard goal into simpler pieces reduces total cost even though the precise growth rate differs from $b^L$.
In our running example, the root goal requires over 25 tactic steps across five interacting definitions; after decomposition into four sublemmas, each is solvable in 1--5 steps, reducing the effective complexity substantially.
Moreover, because the $r$ subgoals are logically independent, they can be solved in parallel, further boosting efficiency.

This analysis also yields a concrete design principle: the system should prefer decompositions that produce subgoals of \emph{roughly equal and individually low difficulty}.
The remainder of this section describes how \tool estimates sublemma difficulty from proof-state features and uses this estimate to rank candidate decompositions.

\subsection{Two Dimensions of Proof Difficulty}
\label{sec:features-motivation}

In program verification, each theorem or sublemma asserts that a \emph{program} (or program fragment) satisfies a \emph{specification} (or program property).
The proof must bridge these two worlds: it unfolds program definitions to expose computational behavior, and it manipulates logical formulas to match the specification.
Accordingly, the difficulty of automating such a proof is governed by two complementary dimensions:

\begin{itemize}[leftmargin=*, itemsep=0.3ex]
\item \emph{Program-side complexity.}
  Proving a property about a program requires unfolding its definitions until the relevant computational structure is exposed.
  Sublemmas that reference deeply nested recursive functions, multiple mutually dependent definitions, or complex pattern-matching induce larger proof search spaces.
  Each unfolding step introduces new terms, case splits, and induction hypotheses that the prover must navigate.
  Lightweight indicators of this dimension include surface occurrences of \texttt{match}/\texttt{fix}/\texttt{let}, map/set identifiers, and the size of the context that appears after introductions.

\item \emph{Specification-side complexity.}
  The specification is ultimately a logical formula, typically in first-order logic.
  Its structural complexity directly affects proof search: quantifier alternations require Skolemization and increase the clause set that ATPs must explore; disjunctive and conjunctive branching multiplies the number of proof obligations; and a deeper AST means more rewriting steps before the goal can be closed.
  Lightweight indicators of this dimension include the size of the normalized goal, residual quantifiers, implications, logical connectives, and comparison operators.
\end{itemize}

This two-dimensional view explains why neither dimension alone suffices.
A sublemma with a simple specification but deeply nested program definitions (e.g., a straightforward equality about a recursive function with five nested calls) can be hard because unfolding generates a large search tree.
Conversely, a sublemma involving a trivial program but a complex specification (e.g., a quantified invariant with multiple disjunctive branches) can be hard because the logical structure demands extensive case analysis.
The difficulty model must account for both.

\subsection{Normalized Proof States as Features}
\label{sec:features-extraction}

To estimate difficulty along these two dimensions, we extract features from each candidate sublemma.
A direct approach is text-level analysis, i.e., extracting features from the raw lemma statement, but this is brittle: syntactically similar statements can have vastly different proof complexities depending on how definitions unfold.
Instead, \tool uses \serapi to step into the proof of each sub-lemma $\ell$ and executes a short prefix (\texttt{Proof. repeat intro.}) to normalize the goal into an ``intros-state,'' in which all top-level universal quantifiers have been moved into the context.
We then extract a feature vector $\phi(\ell)$ from this normalized state, combining two groups of features:

\begin{itemize}[leftmargin=*, itemsep=0.3ex]
    \item \emph{Intros-state features} (19 features) are computed from the proof state after \texttt{repeat intro}. They capture the normalized goal, context, and proof-state shape that the automation backend will face;
    \item \emph{Statement features} (9 features) are computed from the raw lemma statement before entering the proof. They preserve declaration-level structure that \texttt{repeat intro} may move into the context or otherwise redistribute.
\end{itemize}

Table~\ref{tab:features} lists all 28 features.
The normalization step is crucial: by executing \texttt{repeat intro}, the intros-state features reflect the \emph{actual} proof state that the automation backend will operate on, rather than the surface syntax of the lemma statement.
This makes them robust to cosmetic differences such as variable naming or quantifier ordering.
The statement features complement the intros-state features by preserving raw counts of binders, implications, connectives, comparisons, and surface \texttt{match}/\texttt{fix}/\texttt{let} constructs before normalization.

\begin{table}[t]
\caption{Complete feature set ($|\phi| = 28$). ``Intros-state'' features are extracted after executing \texttt{repeat intro}; ``Statement'' features are extracted from the raw lemma declaration.}
\label{tab:features}
\centering
\setlength{\tabcolsep}{2.5pt}
\scriptsize
\begin{tabular}{@{}llp{4.2cm}@{}}
\toprule
\textbf{Group} & \textbf{Feature} & \textbf{Description} \\
\midrule
\multicolumn{3}{@{}l}{\emph{Intros-state: proof-state shape}} \\
& \texttt{num\_goals}       & Number of remaining subgoals \\
\addlinespace
\multicolumn{3}{@{}l}{\emph{Intros-state: goal features}} \\
& \texttt{goal\_len}        & Character length of the goal \\
& \texttt{goal\_tok\_count}  & Token count of the goal \\
& \texttt{goals\_total\_len} & Total char length across all goals \\
& \texttt{goals\_max\_len}   & Max char length among all goals \\
& \texttt{forall\_left}      & Residual $\forall$ quantifiers in goal \\
& \texttt{goal\_arrow}       & Implication arrows ($\to$) in goal \\
& \texttt{goal\_logic\_ops}  & Logic connectives ($\wedge, \vee, \neg$) in goal \\
& \texttt{cmp\_ops}          & Comparison operators ($=$, $<$, $\leq$, etc.) \\
& \texttt{is\_contra\_goal}  & Goal is \texttt{False} or a negation (binary) \\
\addlinespace
\multicolumn{3}{@{}l}{\emph{Intros-state: context features}} \\
& \texttt{num\_hyps}         & Number of hypotheses \\
& \texttt{hyp\_total\_len}   & Total char length of all hypotheses \\
& \texttt{hyp\_len\_avg}     & Average hypothesis length \\
& \texttt{hyp\_len\_max}     & Maximum hypothesis length \\
& \texttt{hyp\_tok\_count\_total} & Total token count of hypotheses \\
& \texttt{hyp\_tok\_count\_max}   & Max token count among hypotheses \\
& \texttt{hyp\_logic\_ops\_total} & Logic connectives across hypotheses \\
& \texttt{match\_fix\_let}   & \texttt{match}/\texttt{fix}/\texttt{let} keywords \\
& \texttt{mapset\_tokens}    & Map/set identifiers (e.g., \texttt{find}, \texttt{mem}) \\
\addlinespace
\multicolumn{3}{@{}l}{\emph{Statement: raw declaration features}} \\
& \texttt{stmt\_len}         & Character length of statement \\
& \texttt{stmt\_tok\_count}  & Token count of statement \\
& \texttt{stmt\_forall}      & $\forall$ quantifiers in statement \\
& \texttt{stmt\_exists}      & $\exists$ quantifiers in statement \\
& \texttt{stmt\_arrow}       & Implication arrows in statement \\
& \texttt{stmt\_logic\_ops}  & Logic connectives in statement \\
& \texttt{stmt\_cmp\_ops}    & Comparison operators in statement \\
& \texttt{stmt\_match\_fix\_let} & \texttt{match}/\texttt{fix}/\texttt{let} in statement \\
& \texttt{stmt\_is\_eq\_goal}    & Statement contains $=$ or $\neq$ (binary) \\
\bottomrule
\end{tabular}
\end{table}

\subsection{Difficulty Model and Candidate Ranking}
\label{sec:rank}

The difficulty score is computed using a linear model parameterized by $\theta$:
$ d_\theta(\ell) = \theta^\top \phi(\ell) + \beta $.
The weights $\theta$ are learned offline from execution traces via a pairwise ranking objective (described below), so that the ranker reflects the \emph{actual} success behavior of the underlying automation backend.
We choose a linear model for two reasons.
First, the feature space has 28 dimensions (Table~\ref{tab:features}), and candidate scoring must be cheap enough to run before recursive search.
Second, a linear model is fully interpretable: in our learned weights, goal size and context size receive large positive coefficients, matching the intuition that both the normalized goal and its surrounding hypotheses contribute to automation difficulty.

To rank the full candidate $c$, we aggregate the difficulties of its constituent sublemmas into a scalar, i.e., 
$ \mathrm{Score}(c) = \mathrm{Agg}(\{ d_\theta(\ell) \mid \ell \in L(c) \}) $
where $\mathrm{Agg}()$ is a configurable function ($\max$, $\mathrm{mean}$, or $\mathrm{sum}$).
By default, we use $\max$, reflecting a conservative strategy that penalizes decompositions containing any single ``bottleneck'' sub-lemma.
The valid set $V$ is sorted by ascending $\mathrm{Score}(c)$ (easier first) to form $V_{\text{ranked}}$.
This aggregation directly reflects the cost analysis above: since overall search cost can often be dominated by the hardest sublemma, ranking by $\max$ difficulty preferentially selects decompositions where even the hardest piece remains within the automation's reach.
In our running example (Figure~\ref{fig:decomposition-example}), the winning decomposition has $\max$ difficulty 0.51, while a rejected alternative scored 0.89 due to a sublemma requiring complex inductive reasoning beyond \coqhammer's reach.

Here, ``easier'' means more tractable for the backend, not logically weaker.
A strengthened auxiliary lemma may be logically stronger than the fact needed at the parent goal, yet still receive a favorable score if its normalized proof state matches patterns that \coqhammer can discharge.
\tool therefore can use strengthened lemmas when the LLM proposes them; the ranker only estimates whether the resulting obligations are likely to be executable under the current backend.
The ranker deliberately avoids expensive semantic analysis such as recursive unfolding.
This keeps candidate scoring cheap relative to hammer invocations, but it can miss difficulty hidden behind opaque definitions or relations.

Note that the Rocq-side validity filter alone is not sufficient: validity ensures the parent goal \emph{could} close assuming the sublemmas, but ranking ensures the sublemmas are actually \emph{executable} under the available automation.

\subsection{Offline Trace Collection and Learning}
\label{sec:offline}

The difficulty model described above requires learned weights $\theta$ that reflect the automation backend's actual solving capability.
However, evaluating ranking policies online is expensive and noisy due to stochastic LLM sampling, making it hard to attribute improvements to ranking choices rather than generation variance.
\tool addresses this by strictly separating online execution from offline policy learning.
During online search, the system records a deterministic, reproducible trace comprising:
\begin{itemize}[leftmargin=*, itemsep=0ex]
    \item \texttt{nodes.jsonl}: Goal statements, depths, and intros-state snapshots;
    \item \texttt{candidates.jsonl}: Generated candidates, Rocq verifier errors, difficulty scores, and the binary \texttt{outcome} of whether the recursive attempt succeeded.
\end{itemize}

This frozen trace enables \emph{offline replay}: we can evaluate arbitrary ranking policies (e.g., swapping $\max$ with $\mathrm{sum}$, or changing $\theta$) by re-sorting the recorded candidates and computing end-to-end metrics without issuing new LLM requests.
Because the traces are collected in dense-supervision mode (Section~\ref{sec:solve}), every candidate at every node has a recorded recursive outcome, so re-ranking produces valid end-to-end results regardless of the original execution order.
By decoupling generation from policy evaluation, we can sweep ranking functions and attribute improvements to ranking choices rather than generation variance.

We learn the weights $\theta$ from the recorded traces using a pairwise ranking objective: for each node where at least one candidate succeeded and one failed, we minimize a margin-based loss encouraging the model to assign lower difficulty to successful candidates.
Because the candidate score is built from sublemma-level difficulties aggregated via $\mathrm{Agg}()$, this training implicitly calibrates the notion of difficulty to the capabilities and timeouts of the underlying automation backend.
The difficulty model is trained on a separate dataset of 200 goals randomly sampled from multiple projects in the \coqgym corpus~\citep{Yang19}, with project-level disjointness from all three evaluation benchmarks: none of the training projects overlap with \coqgymtest, \wigdersontest, or \cloverbenchtest.
This separation tests whether the learned weights capture automation-solvability patterns that transfer beyond the training projects.
Algorithm~\ref{alg:offline} summarizes the offline learning pipeline.

\begin{algorithm}[ht]
\caption{Offline Trace Replay and Weight Learning.}
\label{alg:offline}
\small
\begin{algorithmic}[1]
\Require Trace files $\mathcal{T} = \{(\text{nodes}, \text{candidates})\}$ from online runs; feature extractor $\phi$; aggregation function $\mathrm{Agg}$; regularization $\lambda$
\Ensure Learned weights $\theta$

\State \textbf{// Phase 1: Collect pairwise training data}
\State $\mathcal{P} \gets \emptyset$ \Comment{set of (winner, loser) score pairs}
\For{each node $n$ in $\mathcal{T}$ with candidates $\{c_1, \dots, c_m\}$}
  \For{each pair $(c_i, c_j)$ where $c_i$ succeeded and $c_j$ failed}
    \State $s_i \gets \max_{\ell \in L(c_i)} \theta^\top \phi(\ell)$ \Comment{score of hardest sublemma in $c_i$}
    \State $s_j \gets \max_{\ell \in L(c_j)} \theta^\top \phi(\ell)$ \Comment{score of hardest sublemma in $c_j$}
    \State $\mathcal{P} \gets \mathcal{P} \cup \{(s_i, s_j)\}$ \Comment{winner ($c_i$) should score lower (easier)}
  \EndFor
\EndFor

\State \textbf{// Phase 2: Learn weights via margin ranking loss}
\State Solve $\theta \gets \arg\min_\theta \sum_{(s^+, s^-) \in \mathcal{P}} \max\!\big(0,\; \mu - (s^- - s^+)\big)^2 + \lambda \|\theta\|^2$

\State \textbf{// Phase 3: Offline replay evaluation}
\For{each candidate ranking policy $\pi$ to evaluate}
  \State Re-sort candidates in $\mathcal{T}$ using $\pi(\theta)$
  \State Compute end-to-end success rate without new LLM calls
\EndFor
\end{algorithmic}
\end{algorithm}

\section{Evaluation}
\label{sec:eval}

In this section, we present the experimental results. Our experiments are designed to answer the following research questions:
\begin{enumerate}
  \item[\textbf{RQ1}:] How effective is \tool compared to state-of-the-art proof synthesis methods?
  \item[\textbf{RQ2}:] How efficient is \tool in terms of proving time and LLM cost?
  \item[\textbf{RQ3}:] How does each component (difficulty-based ranking, \coqhammer integration) contribute to \tool's performance?
  \item[\textbf{RQ4}:] How sensitive is \tool to its budget and LLM backend, and what gaps remain?
  \item[\textbf{RQ5}:] How do the learned difficulty model and its design choices affect ranking quality?
\end{enumerate}

\subsection{Experimental Setup}
\label{sec:setup}


\subsubsection{Benchmarks.}
We evaluate on three benchmarks.
\coqgymtest and \wigdersontest are widely used in prior work on automated Rocq proof synthesis~\citep{Yang19,kasibatla2026cobblestone,Sanchez20,Blaauwbroek20}:
\coqgymtest contains 100 theorems randomly sampled from diverse \coqgym~\citep{Yang19} projects, while \wigdersontest contains 100 theorems from the \wigderson software-verification project~\citep{Phipathananunth23Wigderson}, following the benchmark selection used by \baseline~\citep{kasibatla2026cobblestone}.
We use Rocq 8.10 for \coqgymtest and Rocq 8.13 for \wigdersontest, matching each benchmark's original environment.

We further introduce \cloverbenchtest, a benchmark of 58 verification problems translated from Rust/Verus to Rocq, drawn from the benchmark suites of AutoVerus~\citep{AutoVerus2025} and VeruSAGE~\citep{VeruSAGE2025}.
Each problem was translated via an automated pipeline, refined with LLM assistance, and manually reviewed by domain experts; all translated specifications compile under the Rocq type-checker.
\cloverbenchtest is substantially harder than the other two benchmarks: the problems involve richer specifications (e.g., sorting invariants, recursive data structures) and are absent from known public Rocq proof corpora, reducing data-leakage concerns.
We release \cloverbenchtest as part of our artifact and use Rocq 8.20 for evaluation.

\subsubsection{Baselines.}
We compare against the following seven systems:
\begin{itemize}[leftmargin=1em,itemsep=0.3ex]
  \item {\coqhammer}~\citep{Czajka18}: the standalone automation that invokes external ATPs/SMT solvers and reconstructs Rocq-checked proofs.
  \item {Proverbot9001}~\citep{Sanchez20}: a neural tactic predictor trained on Rocq proof traces, combined with proof search.
  \item {Tactician}~\citep{Blaauwbroek20}: a lightweight online learning system that predicts tactics from the current proof state.
  \item {ChainOfThought}~\citep{wei2022chain}: an LLM-based baseline that generates complete proof scripts in a single pass with chain-of-thought prompting.
  \item {PALM}~\citep{lu2024proofautomationlargelanguage}: a retrieval-augmented LLM prover combining premise retrieval with proof generation.
  \item {Rango}~\citep{thompson2025rango}: an LLM-based prover with repair and backtracking capabilities.
  \item {\baseline}~\citep{kasibatla2026cobblestone}: an LLM-based proof-repair system that generates whole proofs, localizes errors via fail-safe execution, and recursively retries on failing subproofs.
\end{itemize}
All baselines are re-run under our Rocq toolchain and evaluation budget.

\subsubsection{Evaluation metrics.}
We report \emph{success rate} (fraction of theorems proven).
To quantify complementarity, we report \emph{added value} following \baseline:
$\mathrm{AV}(X\mid Y) \triangleq {|\mathcal{S}_X \setminus \mathcal{S}_Y|}/{|\mathcal{S}_Y|}$,
where $\mathcal{S}_X$ is the set of theorems proved by $X$.
We also report LLM token usage and request counts.


\subsubsection{Implementation.}
We implement \tool in Python on top of \serapi.
The implementation has three key components.
{(1) Recursive decomposition runner.}
The runner maintains a proof context of already proven lemmas and orchestrates the Generate--Rank--Solve loop.
It supports depth limits on recursion, an invocation budget for LLM calls, and trace logging for offline evaluation.
{(2) Difficulty ranking.}
The difficulty model is implemented as a combination of statement heuristics and intros-state features.
The latter requires running a small prefix (e.g., \texttt{repeat intro.}) to stabilize the goal shape before feature extraction.
We support loading learned weights from a JSON model.
{(3) Hammer-first proof checking.}
\tool uses \coqhammer as a fast path at each recursive node.
When \coqhammer succeeds, we record the reconstructed proof script and avoid further LLM sampling for that subgoal.
This design keeps the search efficient by focusing LLM effort on proposing decompositions, while relying on the theorem prover to discharge many of the resulting local obligations.

All main comparisons (including the baselines that require LLM calls) use \gptfive~\citep{OpenAIGPT52} as the shared LLM backend, so differences in Table~\ref{tab:rq1} reflect the proof-synthesis framework rather than model capability.
We additionally report backend-sensitivity experiments with stronger and alternative LLMs in Section~\ref{sec:rq4}; these runs are not used as the basis for the main baseline comparison.
All backends use the same prompt template (included in our artifact) with temperature $= 1.0$ (following \baseline), top-$p$ at the provider default, and max output tokens $= 4096$.
To ensure a fair comparison, we impose a uniform 10-minute wall-clock budget per theorem, with a 30-second per-goal timeout for \coqhammer and a 90-second timeout for LLM requests.
\tool's per-theorem LLM budget (60 requests) matches \baseline's effective budget (10 nodes $\times$ 6 calls per node).
For \tool, unless otherwise stated, we use rollout budget $k{=}8$, top-$B{=}1$, and a maximum recursion depth of 5, with a hard global cap of 60 LLM requests per theorem.
The difficulty model weights $\theta$ are trained on 200 goals from \coqgym projects disjoint from all three evaluation benchmarks (Section~\ref{sec:offline}).

\subsection{RQ1: Effectiveness}
\label{sec:rq1}

Table~\ref{tab:rq1} summarizes overall success rates and \tool's added value relative to baselines.
The results demonstrate that \tool achieves success rates of 55\%, 52\%, and 16\% on \coqgymtest, \wigdersontest, and \cloverbenchtest respectively, outperforming all baselines on every benchmark.
On \cloverbenchtest, this corresponds to 9 proved theorems out of 58, rounded to 16\%.
The strongest baseline is PALM on \coqgymtest (48\%) and \wigdersontest (39\%); \tool improves over PALM by +7, +13, and +13 percentage points.
\baseline, the state-of-the-art proof-repair system, achieves 38\%, 35\%, and 2\% under the same evaluation conditions.

\mypara{Performance on \cloverbenchtest.}
All systems perform substantially worse on \cloverbenchtest, and the gap is not unique to \tool: non-LLM baselines (\coqhammer, Proverbot9001, Tactician) prove 0 theorems, and the strongest LLM baselines (PALM, \baseline) prove only 2--3\%.
This across-the-board drop reflects two properties of \cloverbenchtest that are by design (Section~\ref{sec:setup}).
First, the problems involve richer specifications (e.g., sorting invariants, recursive data structures) translated from Rust/Verus, resulting in deeper proof obligations where even individual subgoals after decomposition frequently lie beyond \coqhammer's reach.
Second, because these problems are absent from known public Rocq proof corpora, systems are less likely to rely on memorized proof patterns and must reason from the translated specifications.
These two factors compound: when the leaf solver cannot discharge the subgoals produced by decomposition, the planning layer loses its main execution mechanism.
Despite this, \tool's +13 percentage point improvement over the strongest baseline on \cloverbenchtest is tied for the largest margin across all three benchmarks (vs.\ +7 on \coqgymtest and +13 on \wigdersontest), indicating that difficulty-aware planning provides strong relative benefit in the out-of-distribution setting where prior approaches struggle.

\begin{table}[t]
\centering
\caption{Success rate (\%) and added value (AV) on three benchmarks. \tool outperforms all baselines, with average AV ranging from 33.7\% to $\infty$. Percentages on \cloverbenchtest are rounded over 58 theorems.}
\label{tab:rq1}
\small
\setlength{\tabcolsep}{3pt}
\begin{tabular}{@{}lcc|cc|cc|cc@{}}
\toprule
 & \multicolumn{2}{c|}{\coqgymtest} & \multicolumn{2}{c|}{\wigdersontest} & \multicolumn{2}{c|}{\cloverbenchtest} & \multicolumn{2}{c}{Total}\\
 & succ.\% & AV & succ.\% & AV & succ.\% & AV & succ.\% & AV \\
\midrule
\coqhammer & 19 & 189.5 & 22 & 136.4 & 0 & $\infty$ & 15.9 & 182.9 \\
Proverbot9001 & 13 & 323.1 & 10 & 420.0 & 0 & $\infty$ & 8.9 & 404.3 \\
Tactician & N/A & N/A & 13 & 300.0 & 0 & $\infty$ & N/A & N/A \\
ChainOfThought & 45 & 37.8 & 31 & 67.7 & 2 & 800.0 & 29.8 & 59.7 \\
PALM & 48 & 27.1 & 39 & 41.0 & 3 & 350.0 & 34.5 & 33.7 \\
Rango & N/A & N/A & N/A & N/A & 2 & 800.0 & N/A & N/A \\
\baseline & 38 & 78.9 & 35 & 57.1 & 2 & 800.0 & 28.7 & 78.4 \\
\midrule
\tool & \textbf{55} &   & \textbf{52} &   & \textbf{16} &   & \textbf{45.0} &   \\
\bottomrule
\end{tabular}
\end{table}

Since \tool's framework is modular, we can replace \coqhammer with \baseline as the leaf-goal solver.
Because combining two LLM-based systems increases per-theorem cost, we extend the wall-clock budget to 20 minutes for this configuration.
Across all three benchmarks (258 theorems), \tool with \coqhammer proves 116 theorems, while \tool with \baseline proves 126---an additional 10 theorems.
The additional theorems are distributed as 7 on \coqgymtest, 3 on \wigdersontest, and 0 on \cloverbenchtest.
This result suggests that \baseline, despite being a standalone proof system, functions effectively as a more powerful leaf-goal solver within \tool's framework: like \coqhammer, it discharges individual subgoals, but its LLM-based generation can handle goals beyond first-order ATP.
\tool's planning layer still decomposes hard theorems into subgoals that neither \coqhammer nor \baseline can solve monolithically.

\begin{takeawaybox}
\mypara{Takeaway.}
\tool outperforms all baselines on every benchmark, improving over the strongest baseline by up to +13 percentage points.
Its modular framework further benefits from stronger leaf solvers: replacing \coqhammer with \baseline proves 10 additional theorems.
\end{takeawaybox}

\subsection{RQ2: Efficiency}
\label{sec:rq2}

\begin{figure}[t]
\centering
\includegraphics[width=\columnwidth]{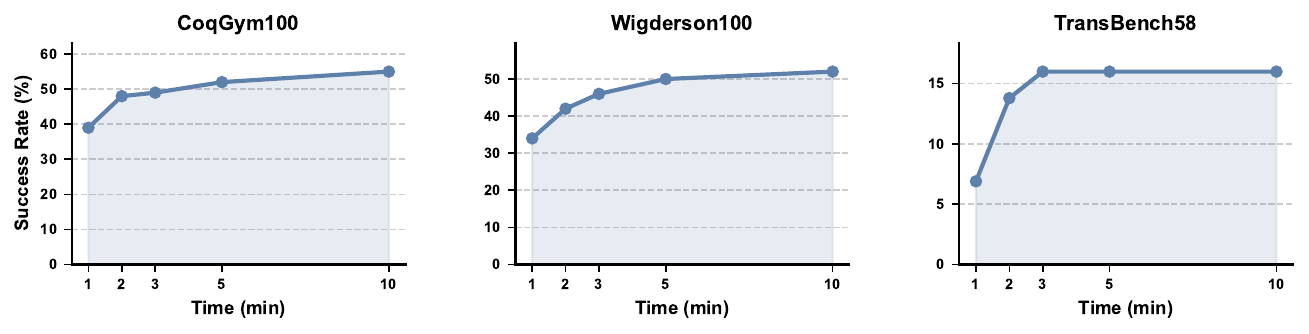}
\caption{Cumulative success rate vs.\ wall-clock time.
Over 90\% of successful proofs complete within 5 minutes; the curve plateaus sharply thereafter, confirming that \tool's hammer-first recursive strategy converges quickly.}
\label{fig:time-curve}
\end{figure}

Figure~\ref{fig:time-curve} shows the cumulative success rate as a function of wall-clock time.
The majority of successful proofs complete quickly: on all three benchmarks, over 90\% of the eventually proved theorems are solved within 5 minutes, and the median solving time is approximately 1--2 minutes.
Beyond the 5-minute mark, almost no new theorems are proved, indicating that \tool's recursive decomposition converges rapidly once a viable proof plan is found.
This efficiency stems from the hammer-first design: \coqhammer typically resolves leaf subgoals in seconds, so the dominant cost per proof tree is the small number of LLM calls at non-leaf nodes rather than repeated generation-and-repair cycles.
In contrast, \baseline's iterative repair loops are expensive: under the same 10-minute budget it achieves only 38\%, 35\%, and 2\%, because many of its proofs require multiple repair iterations that exceed the time limit.

\begin{figure}[ht]
\centering
\begin{subfigure}[b]{0.47\columnwidth}
\centering
\includegraphics[width=\textwidth]{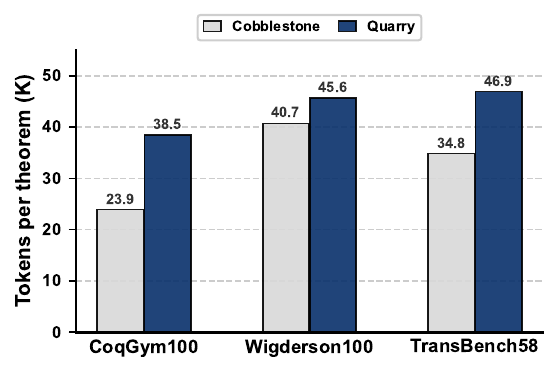}
\caption{Token usage per theorem}
\end{subfigure}
\hfill
\begin{subfigure}[b]{0.47\columnwidth}
\centering
\includegraphics[width=\textwidth]{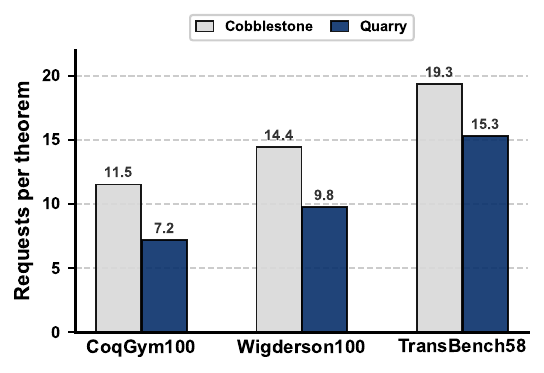}
\caption{Requests per theorem}
\end{subfigure}
\caption{RQ2: Per-theorem LLM cost. \tool uses fewer requests per theorem than \baseline on all benchmarks despite proving more theorems.}
\label{tab:cost}
\end{figure}

Figure~\ref{tab:cost} reports per-theorem token usage and LLM request counts for \tool and \baseline on the three benchmarks.
\tool uses fewer requests per theorem than \baseline on all benchmarks (e.g., 7.2 vs.\ 11.5 on \coqgymtest) while proving more theorems, reflecting the efficiency of difficulty-aware ranking in avoiding fruitless decomposition paths.
On \wigdersontest, \tool's per-theorem token usage is higher (45.6K vs.\ 40.7K) despite fewer requests (9.8 vs.\ 14.4), because its structured decomposition prompts include richer context.

\begin{takeawaybox}
\mypara{Takeaway.}
Over 90\% of \tool's successful proofs complete within 5 minutes, with a median solving time of 1--2 minutes.
\tool also uses fewer LLM requests than \baseline on all benchmarks.
The hammer-first design avoids costly repair loops that dominate \baseline's execution time.
\end{takeawaybox}

\subsection{RQ3: Component Contributions}
\label{sec:rq3}

We ablate the two main design choices---difficulty-based ranking and \coqhammer integration---to understand their individual contributions (Table~\ref{tab:rq2}).

For ranking, we compare \tool against \toolnorank, which keeps candidate verification but executes candidates in the default (unranked) order.
Ranking improves success from 51\% to 55\% on \coqgymtest (+4), from 49\% to 52\% on \wigdersontest (+3), and from 12\% to 16\% on \cloverbenchtest (+4).
\toolnorank also uses more total LLM requests (785 vs.\ 720 on \coqgymtest, 1{,}030 vs.\ 975 on \wigdersontest), because without ranking it more often pursues decomposition paths that fail deep in the recursion tree.
Difficulty-aware ranking thus contributes to both effectiveness and efficiency.

For \coqhammer integration, we compare three configurations: \tool with \coqhammer at every recursive node, \toolnohammer (no hammer calls), and a one-shot union (hammer only on root, then \toolnohammer).
On \coqgymtest, removing \coqhammer reduces success from 55\% to 41\%; a one-shot union recovers to 49\%; invoking \coqhammer at every recursive node adds +6 percentage points over one-shot.
On \wigdersontest, the gap is even more pronounced: 52\% vs.\ 23\% vs.\ 40\%.
This confirms that \coqhammer's value compounds with decomposition depth: as \tool creates deeper proof trees, hammer opportunities at leaf nodes accumulate.

\begin{table}[t]
\centering
\caption{Component ablations (success rate, \%). Both our difficulty-based ranking and CoqHammer integration contribute to the overall performance.}
\label{tab:rq2}
\small
\setlength{\tabcolsep}{3pt}
\begin{tabular}{@{}lccc@{}}
\toprule
\textbf{Configuration} & \textbf{\coqgymtest} & \textbf{\wigdersontest} & \textbf{\cloverbenchtest} \\
\midrule
\multicolumn{4}{@{}l}{\textbf{Ranking ablation}} \\
\toolnorank & 51 & 49 & 12 \\
\tool (with ranking) & \textbf{55} & \textbf{52} & \textbf{16} \\
\addlinespace
\multicolumn{4}{@{}l}{\textbf{Hammer integration ablation}} \\
\coqhammer only & 19 & 22 & 0 \\
\toolnohammer & 41 & 23 & 9 \\
\toolnohammer $\cup$ \coqhammer & 49 & 40 & 9 \\
\tool (hammer at every node) & \textbf{55} & \textbf{52} & \textbf{16} \\
\bottomrule
\end{tabular}
\end{table}

\begin{takeawaybox}
\mypara{Takeaway.}
Both ranking and deep \coqhammer integration are individually significant.
Ranking steers the budget toward hammer-solvable decompositions, while invoking \coqhammer at every node captures leaf-level goals that LLM-only methods miss.
\end{takeawaybox}

\subsection{RQ4: Budget Sensitivity, LLM Backends, and Remaining Gaps}
\label{sec:rq4}

\begin{figure}[t]
\centering
\includegraphics[width=\columnwidth]{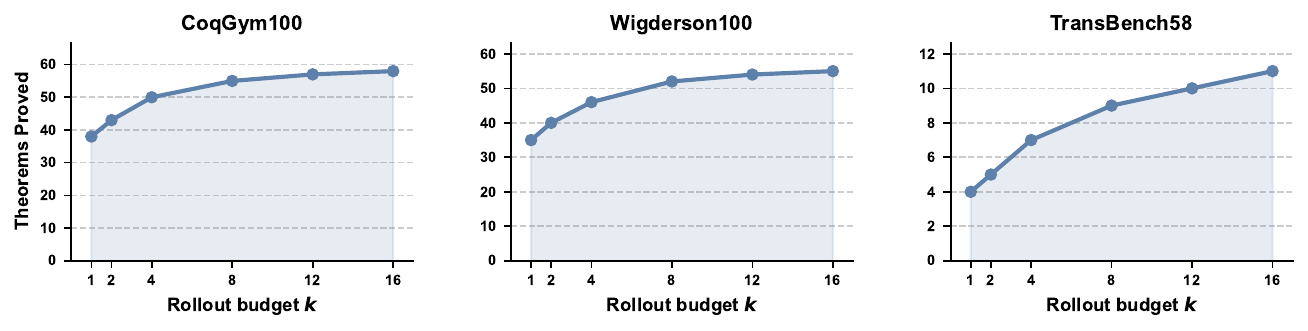}
\caption{Effect of rollout budget $k$ on the number of proved theorems. The default $k{=}8$ captures most of the benefit; gains plateau beyond $k{=}12$.}
\label{fig:rq3-rollout}
\end{figure}

Figure~\ref{fig:rq3-rollout} shows how the number of proved theorems varies with the rollout budget $k$ (candidates sampled per node).
At $k=1$ (single candidate, no selection), \tool proves 38, 35, and 4 theorems on the three benchmarks.
Increasing $k$ improves performance up to $k \approx 12$, after which gains plateau;
the default $k=8$ already captures most of the benefit (55, 52, and 9 theorems) while keeping the per-theorem LLM budget at 60 requests.

\begin{figure}[t]
\centering
\begin{minipage}[t]{0.48\columnwidth}
\centering
\includegraphics[width=\textwidth]{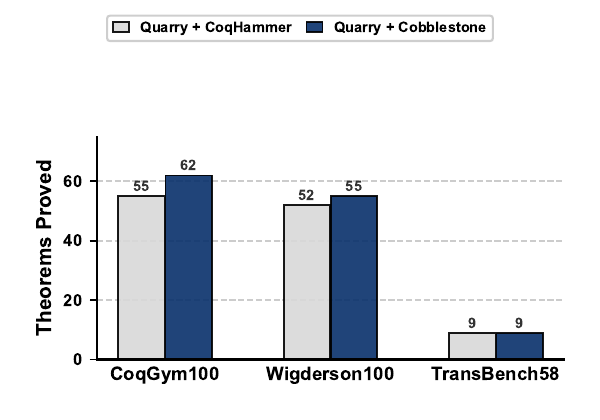}
\captionof{figure}{Effect of leaf-goal solver. Replacing \coqhammer with \baseline at leaf nodes proves additional theorems on \coqgymtest and \wigdersontest.}
\label{tab:complementarity}
\end{minipage}
\hfill
\begin{minipage}[t]{0.48\columnwidth}
\centering
\includegraphics[width=\textwidth]{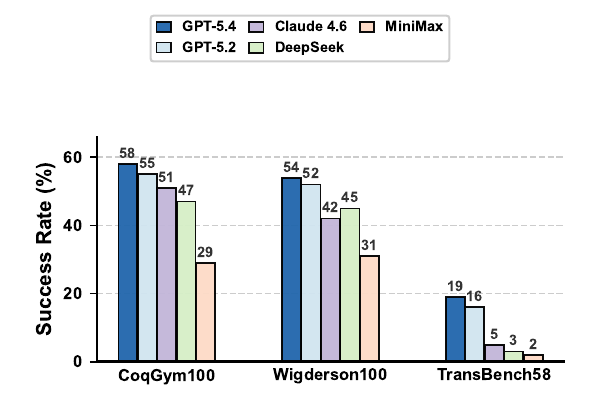}
\captionof{figure}{Sensitivity to LLM backends. The main comparison uses \gptfive; stronger and alternative backends are reported only as sensitivity experiments.}
\label{tab:rq3-models}
\end{minipage}
\end{figure}

Figure~\ref{tab:rq3-models} reports backend-sensitivity results under identical budgets and prompts.
\gptfive is the controlled backend used for the main comparison in Table~\ref{tab:rq1}; \tool improves further with GPT-5.4, reaching 58\%, 54\%, and 19\% on the three benchmarks.
Claude Sonnet 4.6 with thinking reaches 51\%, 42\%, and 5\%.
MiniMax-M2.5~\citep{MiniMaxM25} and DeepSeek-v3.2~\citep{DeepSeekV3.2} provide additional sensitivity points and still outperform standalone \coqhammer on all three benchmarks.
The gap between backends is largest on \cloverbenchtest, suggesting that the harder benchmark amplifies differences in model capability.

The reported main results are single runs.
To estimate run-to-run variability, we repeated the MiniMax-M2.5 configuration two additional times because it was the most cost-efficient backend: the three runs achieved 29/31/28\% on \coqgymtest, 31/33/33\% on \wigdersontest, and 2/0/2\% on \cloverbenchtest, corresponding to standard deviations of at most 1.5 percentage points.
We therefore keep Table~\ref{tab:rq1} as a single-run controlled comparison and use the repeated MiniMax-M2.5 runs only as a cost-conscious estimate of sampling variability under the same search policy.

We also examine the proof trees of the 116 successfully proved theorems (Table~\ref{tab:decomp-stats}).
\tool generates an average of 1.72 sublemmas per theorem, of which 69\% are leaf nodes proved directly by the automation backend.
This shows that a single round of decomposition is usually sufficient to reduce goals to the backend's reach.
\coqgymtest theorems require fewer sublemmas on average (1.40, max 11) than \wigdersontest (2.15, max 25), reflecting the greater structural complexity of the software-verification proofs in \wigdersontest.
The \cloverbenchtest average is lower (1.22) because the solved subset is dominated by tasks where one key translated program invariant suffices; its difficulty lies more in finding that invariant than in building deep decomposition trees.

\begin{table}[t]
\centering
\caption{RQ4: decomposition statistics for 116 successfully proved theorems.}
\label{tab:decomp-stats}
\small
\setlength{\tabcolsep}{3pt}
\begin{tabular}{@{}lcccc@{}}
\toprule
& \textbf{\coqgymtest} & \textbf{\wigdersontest} & \textbf{\cloverbenchtest} & \textbf{Total} \\
\midrule
Proved theorems  & 55   & 52   & 9   & 116 \\
Total nodes      & 132  & 164  & 20  & 316 \\
Total sublemmas  & 77   & 112  & 11  & 200 \\
Avg              & 1.40 & 2.15 & 1.22 & 1.72 \\
Max              & 11   & 25   & 3   & 25 \\
\bottomrule
\end{tabular}
\end{table}

A recurring failure mode is ``internal progress'' without explicit subgoals: proofs that require long sequences of rewriting or unfolding before any case split introduces new goals.
\tool is less susceptible to this than \baseline because it proposes arbitrary sublemmas rather than relying on Rocq's built-in subgoal creation; however, when even the LLM cannot identify a meaningful decomposition, \tool still fails.

\begin{takeawaybox}
\mypara{Takeaway.}
\tool's effectiveness scales with rollout budget up to $k{\approx}12$ and benefits from stronger LLM backends.
The hammer-first recursive strategy is the dominant execution mechanism at leaf nodes.
Inherently sequential proofs without decomposition structure remain the primary failure mode, suggesting that extending the framework to handle ``internal progress'' is a promising direction.
\end{takeawaybox}

\subsection{RQ5: Difficulty Model and Learning Evaluation}
\label{sec:rq5}

The ranking component involves two key design choices: (1)~the feature set used by the difficulty model, and (2)~the aggregation function $\mathrm{Agg}$.
We evaluate both using offline replay (Algorithm~\ref{alg:offline}), which re-sorts the recorded candidates under different configurations without issuing new LLM requests, ensuring that all comparisons reflect ranking quality rather than generation variance.
The difficulty model is trained on 200 goals from \coqgym (disjoint from all test sets) and applied without further tuning to the three benchmarks; \coqgymtest is in-distribution (same corpus, different theorems and projects), while \wigdersontest and \cloverbenchtest are cross-distribution targets.

Our feature set combines 19 intros-state features (extracted after \texttt{repeat intro} via \serapi) with 9 statement-level features (Table~\ref{tab:features}).
We compare three configurations via offline replay: statement features only, intros-state features only, and the full 28-feature set (Table~\ref{tab:rq4}, top).
Statement-only features (9 feat.) achieve 49\%, 51\%, and 9\%---actually \emph{below} \toolnorank on \coqgymtest (51\%) and \cloverbenchtest (12\%), indicating that coarse-grained textual statistics suffer from distribution shift when the target domain diverges from the training set.
Intros-state features (19 feat.) achieve 54\%, 50\%, and 14\%, substantially outperforming statement-only on \cloverbenchtest because normalized proof-state structure provides more robust cross-distribution signals than raw syntax.
The full 28-feature set achieves the best results on all three benchmarks (55\%, 52\%, 16\%): the two feature groups are complementary, as intros-state features provide normalized proof-state signals while statement features preserve raw declaration structure before \texttt{repeat intro} moves or reshapes it into the context.

Table~\ref{tab:rq4} (bottom) compares three aggregation functions.
$\max$ achieves the highest success rate on all benchmarks (55\%, 52\%, 16\%), confirming the bottleneck principle: a decomposition is only as easy as its hardest sublemma.
$\mathrm{Mean}$ performs worst (50\%, 49\%, 10\%) because averaging can mask a single hard sublemma when the rest are easy.

\begin{table}[t]
\centering
\caption{RQ5: effect of feature set and aggregation function on success rate (\%), evaluated via offline replay.
The difficulty model is trained on 200 \coqgym goals (project-level disjoint from all test sets).
\toolnorank (no ranking) achieves 51 / 49 / 12 as reference.}
\label{tab:rq4}
\small
\setlength{\tabcolsep}{3pt}
\begin{tabular}{@{}lccc@{}}
\toprule
\textbf{Configuration} & \textbf{\coqgymtest} & \textbf{\wigdersontest} & \textbf{\cloverbenchtest} \\
\midrule
\multicolumn{4}{@{}l}{\textbf{Feature set ablation}} \\
Statement only (9 feat.)     & 49 & 51 & 9 \\
Intros-state only (19 feat.) & 54 & 50 & 14 \\
Full (28 feat., default)     & \textbf{55} & \textbf{52} & \textbf{16} \\
\addlinespace
\midrule
\multicolumn{4}{@{}l}{\textbf{Aggregation function}} \\
$\mathrm{sum}$  & 53 & 51 & 14 \\
$\mathrm{mean}$ & 50 & 49 & 10 \\
$\max$ (default) & \textbf{55} & \textbf{52} & \textbf{16} \\
\bottomrule
\end{tabular}
\end{table}

\begin{takeawaybox}
\mypara{Takeaway.}
Proof-state normalization via \serapi is an important design choice: intros-state features provide domain-agnostic difficulty signals that transfer across distributions, while statement features alone can even hurt when the target domain diverges from the training set.
The full feature set is strictly best, as the two groups provide complementary and partially redundant signals.
Among aggregation functions, $\max$ is the most effective, validating the bottleneck principle that ranking should target the hardest sublemma in each candidate.
\end{takeawaybox}

\section{Threats to Validity}
\label{sec:threats}

We discuss three external validity threats and the inherent limitations of \tool.

\mypara{Benchmark leakage.}
LLM evaluations may be affected by overlap between benchmarks and model training data.
We mitigate this by reporting multiple datasets with different origins and by including \cloverbenchtest, which contains Rust-to-Rocq translations absent from known public Rocq proof corpora.

\mypara{Model and environment dependence.}
Performance can vary with the LLM backend, prompt template, Rocq version, and ATP configuration.
We report all configurations in our artifact and provide scripts to reproduce our runs.
We report backend sensitivity across five LLM backends (Section~\ref{sec:rq4}) to quantify this dependence.

\mypara{Budget sensitivity.}
Our method is budgeted by recursion depth, rollout count, and LLM samples.
We report success as a function of the rollout budget $k$ (Section~\ref{sec:rq4}) and include per-benchmark cost metrics to enable fair comparison.

\mypara{Limitations.}
\label{sec:limitations}
Beyond the threats above, \tool has several inherent limitations.
First, proofs that make ``internal progress'' without introducing explicit subgoals---e.g., long chains of equational rewrites---lack natural decomposition points, and \tool provides no benefit over monolithic tactics for such sequential obligations (Section~\ref{sec:rq4}).
Second, \tool's pipeline is one-shot: it cannot use execution feedback to generate new candidates when all $k$ pre-generated decompositions fail for the same reason.
Third, the overall ceiling is largely determined by \coqhammer's capabilities, as it discharges the majority of leaf subgoals; goals requiring induction or higher-order reasoning remain difficult even after decomposition.
Finally, \tool's implementation is Rocq-specific (\serapi, \coqhammer); porting to Lean~4 or Isabelle would require reimplementing the proof-state interface and leaf-solver integration.

\section{Related Work}
\label{sec:related}

Several systems discussed below target Isabelle or Lean rather than Rocq; we include them to position our contributions but do not compare experimentally, as cross-prover evaluation involves incompatible proof languages and benchmarks.

\mypara{Automated proof search.}
Hammers such as \coqhammer~\citep{Czajka18} and Sledgehammer~\citep{Paulson23} combine interactive provers with external ATP/SMT solvers and reconstruct kernel-checked proofs.
Premise selection---identifying relevant lemmas to feed to the prover---has been studied extensively using corpus-based, kernel-based, and learned methods~\citep{Alama14,Irving16,Wang17};
more recently, LeanDojo~\citep{Yang23} and Magnushammer~\citep{Mikula24} train models to select premises at each proof step, while Rango~\citep{thompson2025rango} additionally retrieves similar proof scripts.
Thor~\citep{Jiang2022Thor} fine-tunes an LLM to decide when to invoke Sledgehammer versus predicting a tactic, integrating hammers with neural search.
A separate line of work learns to predict tactics directly from proof states: representative systems include GamePad~\citep{Huang19}, CoqGym/ASTactic~\citep{Yang19}, Proverbot9001~\citep{Sanchez20}, Graph2Tac~\citep{Blaauwbroek24ICML}, and LISA~\citep{Jiang21}, spanning architectures from recurrent networks to GNNs and transformer LLMs~\citep{first2020tactok,Ringer21a,bansal2019learning}.
In the Rocq ecosystem, Proverbot9001 outperforms earlier non-LLM systems (e.g., ASTactic, TacTok, Diva, Passport)~\citep{Yang19,first2020tactok,Ringer21a,Sanchez-Stern23toplas}, and Tactician~\citep{Blaauwbroek20} demonstrates that lightweight online learning can be effective.
All of these approaches operate at the level of a \emph{single} goal---selecting premises, predicting one tactic at a time, or deciding whether to invoke a hammer---and do not perform multi-step decomposition or planning.
\tool instead decomposes long-range proof obligations into subproblems, ranks alternative decompositions by estimated solvability, and selects among them under a bounded budget; premise selection methods are complementary and could be integrated into its hammer calls.

\mypara{LLM-based proof planning and decomposition.}
A growing body of work uses LLMs to automate proving; we organize it by how explicitly it plans and evaluates proof structure.
\emph{(1) Whole-proof generation and unplanned search.}
The most direct approaches generate or repair a complete proof and check it post-hoc, without explicitly planning the proof structure.
PALM~\citep{lu2024proofautomationlargelanguage} samples whole proofs and applies retrieval-augmented repair;
Baldur~\citep{First23fse} fine-tunes an LLM to repair whole Isabelle proofs from error messages;
and seL4-prover~\citep{he2026neuro} use a neuro-symbolic \emph{tree search} to generate proofs for systems-software verification.
These methods entangle planning with execution---committing the proof budget to a full script before its structure is known to be sound, or searching without a global plan---so a single wrong turn fails the whole attempt.
\emph{(2) Planning with limited applicability.}
A second line first produces a proof skeleton and then fills its gaps, separating \emph{how to prove} from the proof itself.
Draft-Sketch-Prove~\citep{jiang2023draft} and its variants~\citep{cao2026reviving,varambally2025hilbert} translate an informal proof into a formal sketch whose holes are discharged by Sledgehammer, but it depends on an informal proof that is usually unavailable, and this draft-then-formalize strategy is ineffective for program verification in particular.
POETRY~\citep{wang2024proving} fine-tunes an LLM to predict the proof points at which to recurse, generating the sketch level by level, but relies on model fine-tuning.
\emph{(3) Decomposition with quality estimation.}
A third line explicitly decomposes a goal into sub-lemmas, and some of these works further assess how good a decomposition is.
LEGO~\citep{wanglego} prompts an LLM to emit a complete Isabelle theory of helper lemmas and a main proof, accumulating a skill library across problems, but does not estimate whether a given decomposition is easy to solve.
Closest in spirit to \tool is Goedel-Code-Prover~\citep{li2026goedel}, which likewise searches and \emph{scores} candidate decompositions hierarchically, using an AST-based scoring function together with property-based testing to judge decomposition quality---but this requires the verified artifact to be executable code with generatable tests, restricting it to the code-verification setting.
The most closely related experimental baseline is \baseline~\citep{kasibatla2026cobblestone}, which couples whole-proof generation with recursive repair: it generates whole proofs, localizes errors via fail-safe execution, keeps the working subproofs, and recursively retries the failing ones; however, its decomposition is driven by Rocq's built-in subgoal mechanism (e.g., \texttt{split}, \texttt{induction}), limiting it to subgoals that arise from tactic execution.

\tool differs from these systems in three respects.
First, it plans before it proves: it generates only the decomposition \emph{structure}---sub-lemma statements and a target proof---and validates that structure via Rocq type-checking under temporarily \texttt{Admitted} sub-lemmas, before investing any budget in the sub-lemma proofs.
Second, it can introduce \emph{arbitrary} sub-lemmas via standalone \texttt{Lemma} declarations, not just the subgoals produced by built-in goal-splitting, which directly addresses the ``internal progress'' limitation identified in \baseline's own analysis.
Third, it ranks candidate decompositions with a lightweight difficulty model: unlike Goedel-Code-Prover, which relies on AST-based scoring and property-based testing (and thus on executable code and tests), \tool estimates sub-lemma difficulty from \emph{proof-state} features alone---requiring no execution and applying to general Rocq developments---and tries only the top-ranked candidates under a bounded budget, preventing the failure mode where a recursive prover exhausts its budget on the first plausible but ultimately intractable plan.
Finally, unlike Baldur and POETRY, \tool requires no model fine-tuning, making it straightforward to swap in stronger LLMs as they become available.

\section{Conclusion}
\label{sec:conclusion}

\tool demonstrates that \emph{planning}---actively generating, evaluating, and selecting proof decompositions before execution---is a more effective strategy for automated proof synthesis than reactive step-by-step proving.
By separating the LLM's role (proposing verifiable proof plans) from the prover's role (executing and checking them via \coqhammer), \tool achieves success rates of 55\%, 52\%, and 16\% on three benchmarks, outperforming all baselines while using fewer LLM requests.
Planning-based decomposition, combined with difficulty-aware ranking and deep hammer integration, provides effective levers for scaling automated proof synthesis.

We see two main directions for future work.
First, closing the loop between planning and execution via an \emph{agentic} architecture: a proof agent that observes execution failures and revises its decomposition strategy accordingly, enabling recovery from systematic generation errors and dynamic granularity selection for the ``internal progress'' proofs that the current fixed pipeline cannot handle (Section~\ref{sec:rq4}).
Second, adapting the approach to other ITPs such as Lean~4 or Isabelle to test the generality of the framework beyond the Rocq ecosystem.

\begin{acks}
We thank the anonymous reviewers for their constructive feedback. This work is supported by the Fundamental and Interdisciplinary Disciplines Breakthrough Plan of the Ministry of Education of China (No.\ JYB2025XDXM118), the Frontier Technologies R\&D Program of Jiangsu (BF2024059), the National Natural Science Foundation of China (No. 92582201, 62572229), and the ``111 Center'' (No. B26023). Yuan Yao and Xiaoxing Ma are the corresponding authors.
\end{acks}

\section*{Data-Availability Statement}
Our artifact is available at \url{https://github.com/ningZhang-cs/QUARRY}.
It contains:
(1)~the complete source code of \tool, including the recursive decomposition runner, difficulty ranking module, and \coqhammer integration;
(2)~all three evaluation benchmarks (\coqgymtest, \wigdersontest, and \cloverbenchtest) with their Rocq project environments;
(3)~the learned difficulty model weights;
(4)~prompt templates used for all LLM backends;
and (5)~scripts to reproduce all experiments reported in this paper.

\bibliographystyle{ACM-Reference-Format}
\bibliography{softeng}

\end{document}